\documentstyle[epsf]{mn}		

\title[Peculiar motions of early-type galaxies -- V]
{The peculiar motions of early-type galaxies in two distant regions -- V.
The Mg--{\boldmath$\sigma$} relation, age and metallicity}

\author[Colless et al.]{
{\LARGE \rm Matthew Colless$^1$, David Burstein$^2$, Roger L.\ Davies$^3$, 
Robert K.\ McMahan Jr$^{4,5}$,} \vspace*{6pt} \\ 
{\LARGE \rm R.\ P.\ Saglia$^6$ and Gary Wegner$^7$}  
\vspace*{3pt} \\ 
$^1$Mount Stromlo and Siding Spring Observatories, The Australian 
National University, Weston Creek, ACT 2611, Australia \\
$^2$Department of Physics and Astronomy, Box 871054, Arizona State
University, Tempe, AZ 85287-1504, U.S.A. \\
$^3$Department of Physics, South Road, Durham DH1 3LE, United Kingdom \\
$^4$Dept of Physics and Astronomy, University of North Carolina,
CB 3255 Phillips Hall, Chapel Hill, NC 27599-3255, U.S.A. \\
$^5$P.O. Box 14026, McMahan Research Laboratories, 
79 Alexander Drive, Research Triangle, NC 27709, U.S.A. \\
$^6$Universit\"{a}ts-Sternwarte M\"{u}nchen, Scheinerstra{\ss}e 1, 
D-81679 M\"{u}nchen, Germany \\
$^7$Department of Physics and Astronomy, 6127 Wilder Laboratory,
Dartmouth College, Hanover, NH 03755-3528, U.S.A. }

\date{Accepted ---. Received ---; in original form ---.}

\newlength{\plotwidth}
\newcommand{\etal}{\hbox{et~al.}}
\newcommand{\ie}{\hbox{i.e.}}
\newcommand{\eg}{\hbox{e.g.}}

\newcommand{\cf}{\hbox{cf.}}

\newcommand{\kpc}{\hbox{\,h$_{50}^{-1}$\,kpc}}
\newcommand{\kms}{\hbox{\,km\,s$^{-1}$}}

\newcommand{\mgb}{\hbox{Mg$b$}}
\newcommand{\mgbp}{\hbox{Mg$b^\prime$}}

\newcommand{\mgtwo}{\hbox{Mg$_2$}}

\newcommand{\mgsig}{\hbox{Mg--$\sigma$}}
\newcommand{\mgbpsig}{\hbox{\mgbp--$\sigma$}}
\newcommand{\mgtwosig}{\hbox{\mgtwo--$\sigma$}}

\newcommand{\phpm}{\phantom{$\pm$}}
\newcommand{\gs}
           {\mathrel{\hbox{\rlap{\hbox{\lower4pt\hbox{$\sim$}}}\hbox{$>$}}}}
\newcommand{\ls}
           {\mathrel{\hbox{\rlap{\hbox{\lower4pt\hbox{$\sim$}}}\hbox{$<$}}}}

\newcommand{\plotone}[1]
           {\centering \leavevmode \epsfxsize=\plotwidth \epsfbox{#1}}
\newcommand{\plottwo}[2]
           {\centering \leavevmode \epsfxsize=\plotwidth \epsfbox{#1} 
            \hfill \epsfxsize=\plotwidth \epsfbox{#2}}
\newcommand{\plotfull}[1]
           {\centering \leavevmode \epsfxsize=\textwidth \epsfbox{#1}}

\begin{document}

\maketitle

\begin{abstract}
We have examined the \mgsig\ relation for early-type galaxies in the
EFAR sample and its dependence on cluster properties. A comprehensive
maximum likelihood treatment of the sample selection and measurement
errors gives fits to the global \mgsig\ relation of
$\mgbp=0.131\log\sigma-0.131$ and $\mgtwo=0.257\log\sigma-0.305$. The
slope of these relations is 25\% steeper than that obtained by most
other authors due to the reduced bias of our fitting method. The
intrinsic scatter in the global \mgsig\ relation is estimated to be
0.016~mag in \mgbp\ and 0.023~mag in \mgtwo. The \mgsig\ relation for cD
galaxies has a higher zeropoint than for E and S0 galaxies, implying
that cDs are older and/or more metal-rich than other early-type galaxies
with the same velocity dispersion.

We investigate the variation in the zeropoint of the \mgsig\ relation
between clusters. We find it is consistent with the number of galaxies
observed per cluster and the intrinsic scatter between galaxies in the
global \mgsig\ relation. We find no significant correlation between the
\mgsig\ zeropoint and the cluster velocity dispersion, X-ray luminosity
or X-ray temperature over a wide range in cluster mass. These results
provide constraints for models of the formation of elliptical galaxies.
However the \mgsig\ relation on its own does not place strong limits on
systematic errors in Fundamental Plane distance estimates due to stellar
population differences between clusters.

We compare the intrinsic scatter in the \mgsig\ and Fundamental Plane
(FP) relations with stellar population models in order to constrain the
dispersion in ages, metallicities and $M/L$ ratios for early-type
galaxies at fixed velocity dispersion. We find that variations in age
alone or metallicity alone cannot explain the measured intrinsic scatter
in both \mgsig\ and the FP. We derive the joint constraints on the
dispersion in age and metallicity implied by the scatter in the \mgsig\
and FP relations for a simple Gaussian model. We find upper limits on
the dispersions in age and metallicity at fixed velocity dispersion of
32\% in $\delta t/t$ and 38\% in $\delta Z/Z$ if the variations in age
and metallicity are uncorrelated; only strongly anti-correlated
variations lead to significantly higher upper limits. The joint
distribution of residuals from the \mgsig\ and FP relations is only
marginally consistent with a model having no correlation between age and
metallicity, and is better-matched by a model in which age and
metallicity variations are moderately anti-correlated ($\delta
t/t$$\approx$40\%, $\delta Z/Z$$\approx$50\% and $\rho$$\approx$$-$0.5),
with younger galaxies being more metal-rich.
\end{abstract}
 

\vspace*{2cm}

\section{INTRODUCTION}
\label{sec:intro}

The primary aim of the EFAR project (Wegner \etal\ 1996; Paper~1) is to
use the tight correlations between the global properties of early-type
galaxies embodied in the Fundamental Plane (FP: Djorgovski \& Davis
1987, Dressler \etal\ 1987) to measure relative distances to clusters of
galaxies in order to investigate peculiar motions and the mass
distribution on large scales. However these global relations also
constrain the dynamical properties and evolutionary histories of
early-type galaxies. For example, Renzini \& Ciotti (1993) show that the
tilt of the FP implies a range in mass-to-light ratio $M/L$ among
ellipticals of less than a factor of three, while the low scatter about
the FP implies a scatter in $M/L$ at any location in the plane of less
than 12\%. Similar reasoning has been used to constrain the star
formation history of cluster ellipticals using the colour--magnitude
relation (Bower \etal\ 1992, Kodama \& Arimoto 1997). Recently the FP,
\mgsig\ and colour--magnitude relations have been followed out to higher
redshifts and used to show that the early-type galaxies seen at
$z$$\sim$1 differ from present-day early-type galaxies in a manner
consistent with passive evolutionary effects (van Dokkum \& Franx 1996,
Ziegler \& Bender 1997, Kelson \etal\ 1997, Ellis \etal\ 1997, Kodama \&
Arimoto 1997, Stanford \etal\ 1998, Kodama \etal\ 1998, Bender \etal\
1998, van Dokkum \etal\ 1998).

In this paper we consider the relation between the central velocity
dispersion $\sigma$ and the strength of the magnesium lines at a rest
wavelength of 5174\AA\ for the early-type galaxies in the EFAR sample.
This \mgsig\ relation connects the dynamical properties of galaxy cores
with their stellar populations. The remarkably small scatter about this
relation (Burstein \etal\ 1988, Guzm\'{a}n \etal\ 1992, Bender \etal\
1993, J{\o}rgensen \etal\ 1996, Bender \etal\ 1998), and its
distance-independent nature, make it a potentially useful constraint on
models of the star formation history of early-type galaxies and a test
for environmental variations in the FP (Burstein \etal\ 1988, Bender
\etal\ 1996).

There are, however, some problems with using the \mgsig\ relation for
probing galaxy formation. Two of these problems are apparent from the
stellar population models (\eg\ Worthey 1994, Vazdekis \etal\ 1996):
(i)~both age and metallicity contribute to the Mg linestrengths in
comparable degree, so that a spread in linestrengths could be due to
either a range of ages or a range of metallicities or some combination;
(ii)~the Mg linestrengths are not particularly sensitive indicators---at
fixed metallicity a difference in age of a factor of ten only results in
a change of 0.05-0.1~mag, while at fixed age a difference of 1~dex in
metallicity gives a change of 0.1--0.2~mag. Thus Mg linestrength
measurements must be accurate in order to yield useful constraints on
the ages and metallicities of stellar populations, and the \mgsig\
relation on its own can only supply constraints on combinations of age
and metallicity and not one or the other separately.

Recently Trager (1997) has suggested that the tightness of the \mgsig\
relation may be the result of a `conspiracy', in that there appears to
be an anti-correlation between the ages and metallicities of the stellar
populations in early-type galaxies at fixed mass which acts to reduce
the scatter in the Mg linestrengths. Trager takes the accurate H$\beta$,
Mg and Fe linestrengths from Gonz\'{a}lez (1993) and applies the stellar
population models of Worthey (1994) to derive ages and abundances from
line indices with different dependences on age and metallicity. He finds
that at fixed velocity dispersion the ages and abundances lie in a plane
of almost constant Mg linestrength, leading him to predict little
scatter in the \mgsig\ relation even for large differences in age or
metallicity---a factor of ten in age (from 1.5~Gyr to 15~Gyr) gives a
spread in \mgtwo\ of only 0.01--0.02~mag. This conclusion depends on the
appropriateness of the single stellar population models and requires
confirmation from further high-precision linestrength measurements. It
can also be tested using the high-redshift samples now becoming
available.

In a similar vein, a number of authors (Ferreras \etal\ 1998, Shioya \&
Bekki 1998, Bower \etal\ 1998) have recently re-examined whether the
apparent passive evolution of the colour--magnitude relation out to
$z$$\sim$1 really implies a high redshift for the bulk of the
star-formation in elliptical galaxies. They conclude that in fact such
evolution can be consistent with a rather broad range of ages and
metallicities if the galaxies assembling more recently are on average
more metal-rich than older galaxies of similar luminosity.

As well as studies focussing on the evolution of the galaxy population,
there have also been investigations of possible variations with local
environment. Guzm\'{a}n \etal\ (1992) have suggested that there are
systematic variations in the \mgsig\ relation which affect estimates of
relative distances based on the FP. They report a significant offset in
the zeropoint of the \mgsig\ relation between galaxies in the core of
the Coma cluster and galaxies in the cluster halo. J{\o}rgensen and
co-workers (1996, 1997) examine a sample of 11 clusters and find a weak
correlation between Mg linestrength and local density within the cluster
which is consistent with this result. Similar offsets are claimed
between field and cluster ellipticals by de Carvalho \& Djorgovski
(1992) and J{\o}rgensen (1997), although Burstein \etal\ (1990) find no
evidence of environmental effects. Such systematic differences could
result from different star-formation histories in different density
environments, producing variations in the mass-to-light ratio of the
stellar population. FP distance measurements would then be subject to
environment-dependent systematic errors leading to spurious peculiar
motions. Where data for field and cluster ellipticals come from
different sources, however, the possibility also exists that any
zeropoint differences are due to uncertainties in the relative
calibrations rather than intrinsic environmental differences.

The \mgsig\ relation has thus become an important diagnostic for
determinations of both the star formation history and the peculiar
motions of elliptical galaxies. Here we examine the \mgsig\ relation in
the EFAR sample, which includes more than 500 early-type galaxies drawn
from 84 clusters spanning a wide range of environments. In \S2 we
summarise the relevant properties of the sample and the techniques used
to determine the \mgb\ and \mgtwo\ linestrength indices, the central
velocity dispersions $\sigma$, and the errors in these quantities. We
present the \mgsig\ relation in \S3 and investigate how it varies from
cluster to cluster within our sample, and with cluster velocity
dispersion, X-ray luminosity and X-ray temperature. In \S4 we compare
our results with the predictions of stellar population models in order
to derive constraints on the ages, metallicities and mass-to-light
ratios of early-type galaxies in clusters. In particular, we consider
the constraints on the dispersion in the ages and metallicities from the
intrinsic scatter in the \mgsig\ relation on its own, and in combination
with the intrinsic scatter in the FP. Our conclusions are given in \S5.
 
\section{THE DATA} 
\label{sec:data}

Here we give a short description of our sample and dataset, with
emphasis on the velocity dispersions and line indices used in this
paper. The interested reader can find more detail on the sample
selection in Paper~1 (Wegner \etal\ 1996); on the measurement,
calibration and error estimation procedures for the spectroscopic
parameters in Paper~2 (Wegner \etal\ 1998); and on the structural and
morphological properties of the galaxies in Paper~3 (Saglia \etal\
1997).

\subsection{The sample}
\label{ssec:sample}

The EFAR sample of galaxies comprises 736 mostly early-type galaxies in
84 clusters. These clusters span a range of richnesses and lie in two
regions toward Hercules--Corona Borealis and Perseus--Pisces--Cetus at
distances of between 6000\kms\ and 15000\kms. In addition to this
program sample we have also observed 52 well-known galaxies in Coma,
Virgo and the field in order to provide a calibrating link to previous
studies.

The EFAR galaxies are listed in Table~2 of Paper~1, and comprise an
approximately diameter-limited sample of galaxies larger than about
20~arcsec with the visual appearance of ellipticals. Photometric imaging
(Paper~3) shows that 8\% are cDs, 12\% are pure Es and 49\% are
bulge-dominated E/S0s; thus 69\% of the sample are early-type galaxies,
with the remaining 31\% being spirals or barred galaxies. We have
obtained spectroscopy for 666 program galaxies, measuring redshifts,
velocity dispersions and linestrength indices (Paper~2). We have used
the redshifts we obtained together with literature redshifts for other
galaxies in the clusters in order to assign program galaxies to physical
clusters. We have used the combined redshift data for these physical
clusters to estimate cluster mean redshifts and velocity dispersions.
The early-type galaxies in our sample span a wide range in luminosity,
size and mass: they have absolute magnitudes from $M_R$=$-$24 to $-$18
($\langle M_R \rangle$=$-$21.6; H$_0$=50\,km\,s$^{-1}$\,Mpc$^{-1}$),
effective radii from 1\kpc\ to 70\kpc\ ($\langle R_e \rangle$=9.1\kpc)
and central velocity dispersions from less than 100\kms\ to over
400\kms\ ($\langle \sigma \rangle$=220\kms). The sample is thus
dominated by early-type galaxies with luminosities, sizes and masses
typical of giant ellipticals.

\subsection{The measurements}
\label{ssec:measure}

We summarise here the procedures used in measuring the redshifts,
velocity dispersions and Mg linestrengths; full details are given in
Paper~2.

\begin{figure}
\centering
\plotone{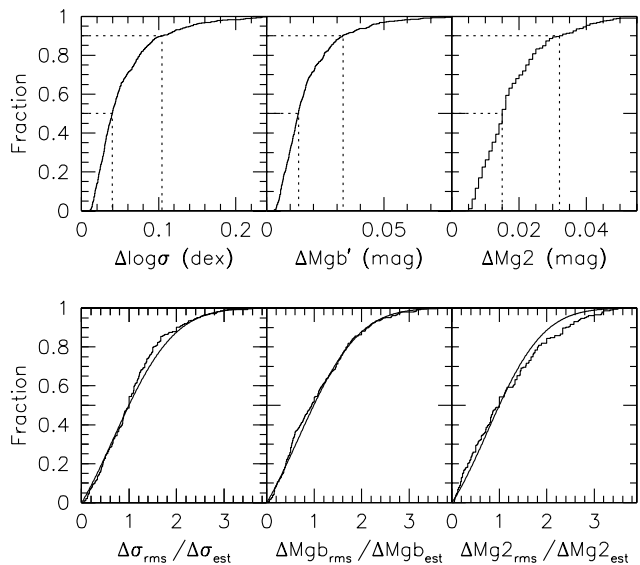}
\caption{A summary of the errors in $\log\sigma$, \mgbp\ and \mgtwo. The
upper panel shows the cumulative distributions of the estimated errors,
with the median and 90th-percentile errors indicated. The lower panel
shows the calibration against the repeat observations: the distribution
of the ratio of rms error to estimated error for objects with repeat
measurements is compared to the predicted distribution assuming the
estimated errors are the true errors.
\label{fig:errsum}}
\end{figure}

Redshifts and velocity dispersions were measured from each observed
galaxy spectrum using the {\sc IRAF} task {\tt fxcor}. Linestrength
indices on the Lick system were determined using the prescription given
by Gonz\'{a}lez (1993). The \mgb\ and \mgtwo\ indices were both
measured: \mgtwo\ because it is the index most commonly measured in
previous work, and \mgb\ because it could be measured for more objects
(as it requires a narrower spectral range) and is better-determined
(being less susceptible to variations in the non-linear continuum
shape). We find it more convenient to express the `atomic' \mgb\ index
in magnitudes like the `molecular' \mgtwo\ index rather than as an
equivalent width in {\AA}ngstroms, since this puts these two indices on
similar footings. The conversion is
\begin{equation}
\mgbp\ = -2.5\log_{10}\left(1-\frac{\mgb}{\Delta\lambda}\right) ~,
\label{eqn:primedef}
\end{equation}
where $\Delta\lambda$ is the index bandpass (32.5\AA\ for \mgb).

Error estimates for each quantity were derived from detailed Monte Carlo
simulations, calibrated by comparisons of the estimated errors with the
results obtained from repeat measurements (over 40\% of our sample had
at least two spectra taken). Two sorts of corrections were applied to
the dispersions and linestrengths: (i)~an aperture correction, based on
that of J{\o}rgensen \etal\ (1995), to account for different effective
apertures sampling different parts of the galaxy profile, and (ii)~a run
correction to remove systematic errors between different observing
setups. After applying these corrections, individual measurements for
each galaxy were combined using a weighting scheme based on the
estimated errors and the overall quality of the spectrum.

\begin{figure*}
\centering
\plottwo{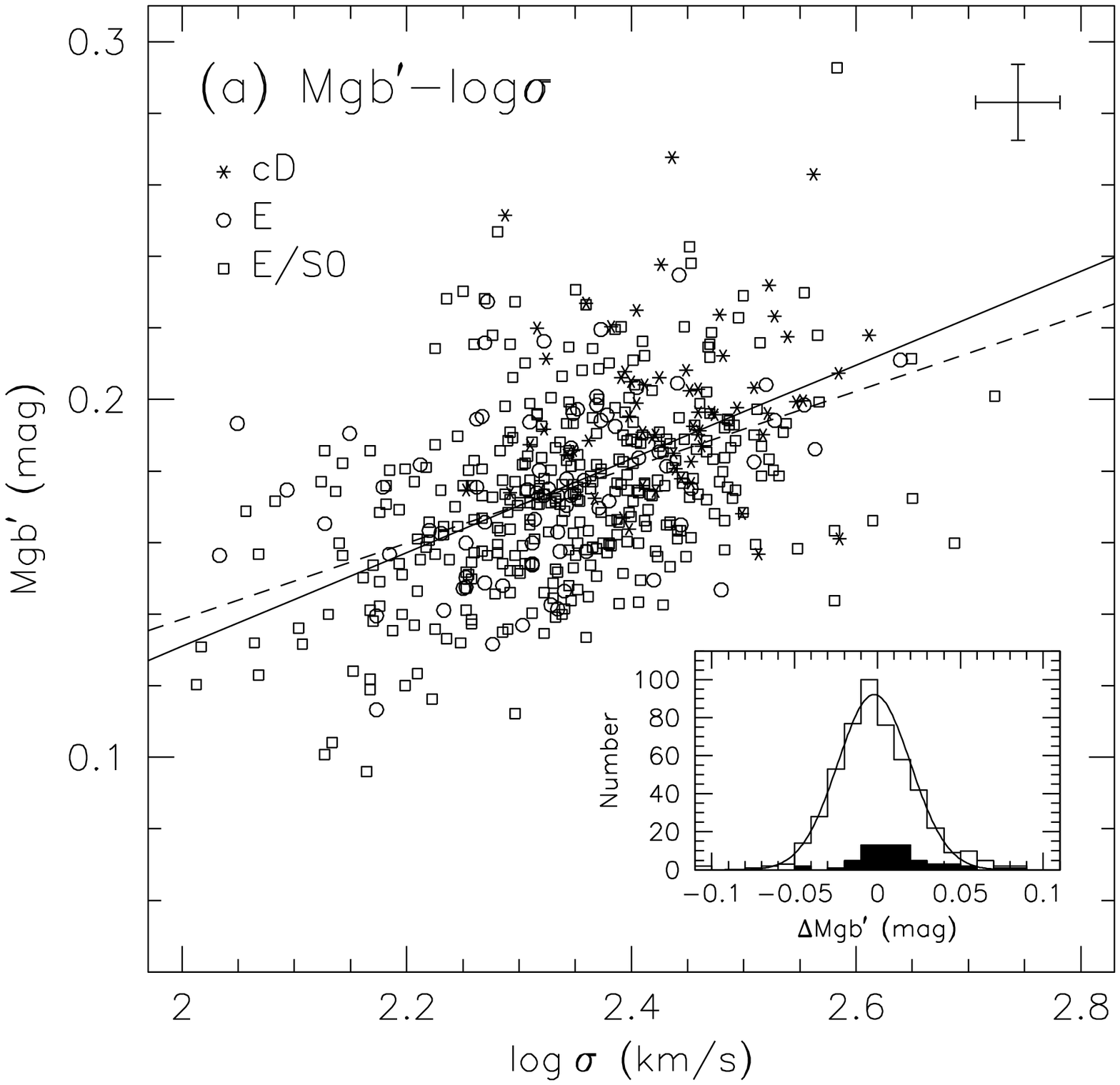}{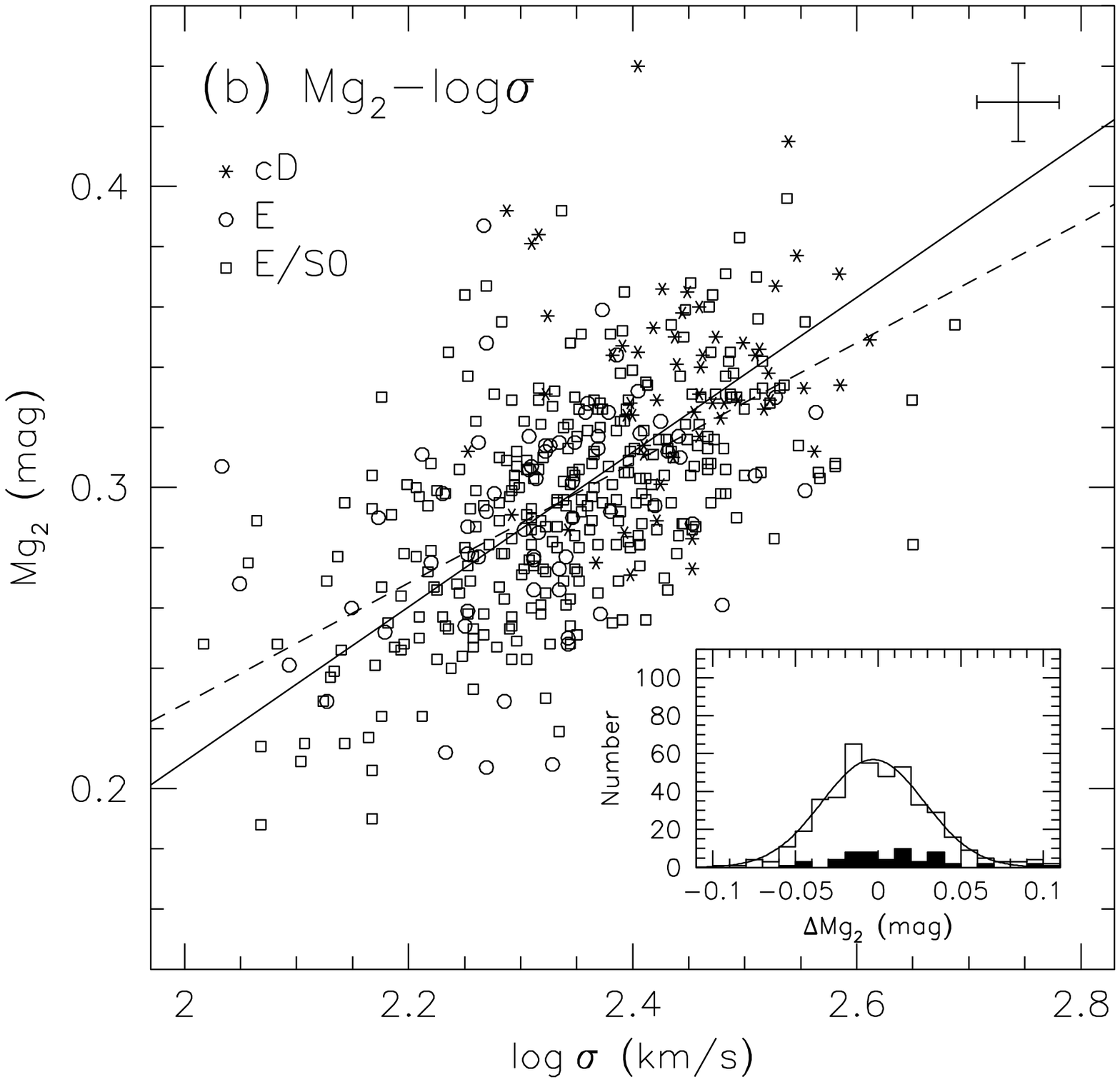}
\caption{The \mgsig\ relations for all early-type galaxies with
linestrength measurements in the EFAR sample: (a)~The \mgbpsig\
relation; (b)~the \mgtwosig\ relation. In both panels E galaxies are
marked as circles, E/S0 galaxies as squares and cD galaxies as
asterisks. Errorbars representing the median errors in each quantity are
shown in the upper right of each panel. The solid line is the ML fit and
the dashed line is the unweighted regression of linestrength on
dispersion. The distributions of residuals in Mg about the ML fit for
all objects are shown in the insets as open histograms. The Gaussians
described by the median residual and the robustly-estimated scatter are
superimposed. The solid histograms are the distributions of residuals
for the cD galaxies, showing their offset from the overall relation.
\label{fig:mgsig}}
\end{figure*}

The median estimated errors in the final combined values are
$\Delta\sigma/\sigma=9.1\%$ (\ie\ $\Delta\log\sigma=0.040$~dex),
$\Delta\mgbp=0.013$~mag and $\Delta\mgtwo=0.015$~mag. The distribution
of estimated errors for each quantity is shown in the upper panel of
Figure~\ref{fig:errsum}. The lower panel of the figure shows how the
error estimates were calibrated against the repeat observations: the
distribution of the ratio of rms error to estimated error for objects
with repeat measurements is compared to the predicted distribution
assuming the estimated errors are the true errors. The initial error
estimates from the simulations have been re-scaled to give the best
match (under a K-S test) to the rms errors from the repeat measurements.
A re-scaling by factors of 0.85 and 1.15 respectively gives good
agreement for the errors in $\sigma$ and \mgb; adding 0.005~mag likewise
gives good agreement for the errors in \mgtwo.

A comparison with the literature (Paper~2, Figure~13) shows that our
dispersions are consistent with previous measurements by Davies \etal\
(1987), Guzm\'{a}n (1993), J{\o}rgensen (1997), Lucey \etal\ (1997) and
Whitmore \etal\ (1985). For the subset of galaxies in common, we
compared our linestrengths with the definitive Lick system measurements
of Trager \etal\ (1998) in order to derive the small zeropoint
corrections required to calibrate our measurements to the Lick system
(Paper~2, Figures~14 \&~15); the overlap of our \mgtwo\ measurements
with those of Lucey \etal\ (1997) also shows consistency (Figure~16,
Paper~2).

\section{THE \mgsig\ RELATION}
\label{sec:mgsig}

\subsection{The global relation} 
\label{ssec:efargals}

In this section we investigate the global \mgsig\ relation found amongst
the entire sample of EFAR galaxies with early-type morphological
classifications (cD, E or E/S0; see definitions in Paper~3) for which we
obtained linestrength measurements. The \mgbpsig\ relation is shown in
Figure~\ref{fig:mgsig}a and the \mgtwosig\ relation in
Figure~\ref{fig:mgsig}b.

\begin{figure}
\centering
\plotone{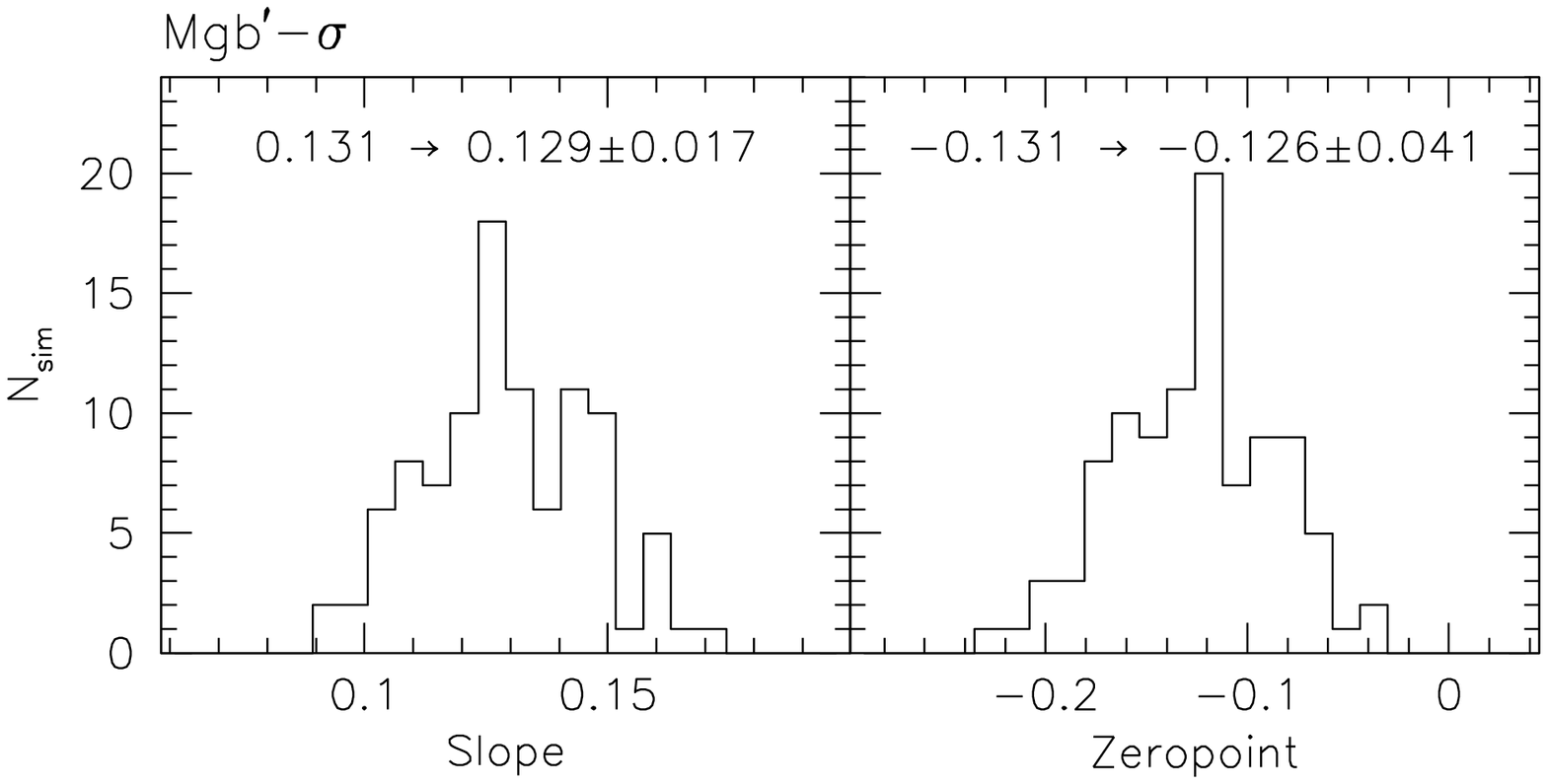} \\
\plotone{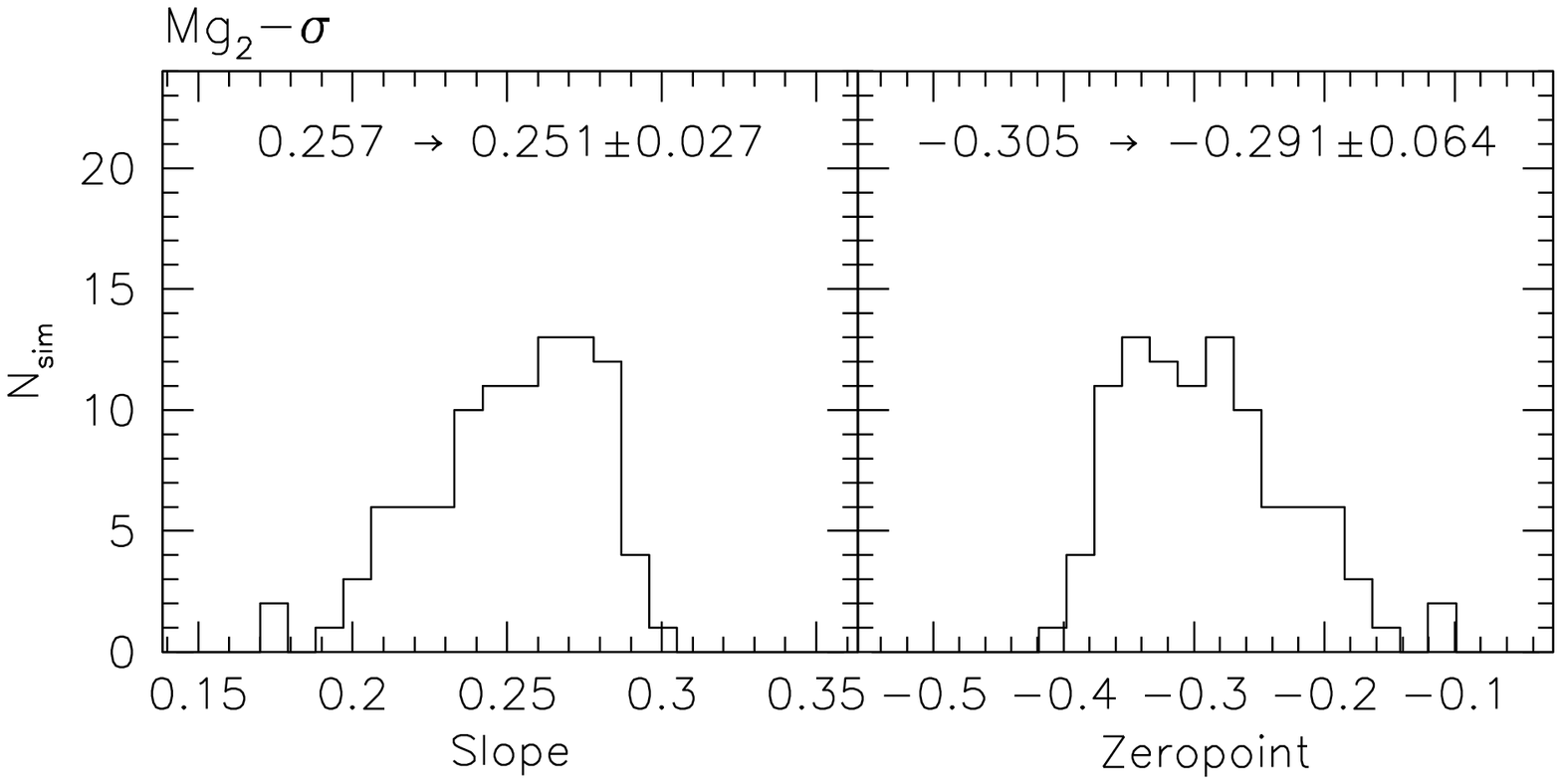}
\caption{Monte Carlo simulations of the ML fits to the \mgsig\
relations. The top panel is for \mgbpsig\ and the bottom panel for
\mgtwosig. In each panel we show the distribution of ML fits to the
slope and zeropoint for a set of 99 simulations of the EFAR dataset. The
input parameters (from the fits to the EFAR data) and the mean and
standard deviation of the fitted parameters (from the simulations) are
given in each case.
\label{fig:mgsigsim}}
\end{figure}

In order to fit linear relations with intrinsic scatter in the presence
of significant measurement errors in both variables, arbitrary censoring
of the dataset and a broad sample selection function, we have developed
a comprehensive maximum likelihood (ML) fitting procedure (Saglia \etal,
in preparation). Excluding galaxies with dispersions less than 100\kms\
or selection probabilities less than 10\%, and also outliers with low
likelihoods, the ML fits to the \mgbpsig\ relation (490 galaxies) and
the \mgtwosig\ relation (423 galaxies) are:
\begin{eqnarray}
\mgbp &=&(0.131\pm0.017)\log\sigma-(0.131\pm0.041) ~, \label{eqn:mlmgbp} \\
\mgtwo&=&(0.257\pm0.027)\log\sigma-(0.305\pm0.064) ~. \label{eqn:mlmgtwo}
\end{eqnarray}
These fits are shown in Figure~\ref{fig:mgsig} as solid lines. The ratio
of the slopes of these relations is consistent with the \mgtwo-\mgbp\
relation we obtained in Paper~2: \mgtwo$\approx$1.94\mgbp$-$0.05. Monte
Carlo simulations of the dataset and fitting process, the results of
which are displayed in Figure~\ref{fig:mgsigsim}, show that there is no
bias in the ML estimates of the slopes and zeropoints, and provide
reliable estimates of the uncertainties in the fit.

The ML fits can be compared to simple regressions of \mgbp\ and \mgtwo\
on $\log\sigma$. These regressions are shown in the figure as dashed
lines, and are:
\begin{eqnarray}
\mgbp  & = & (0.104\pm0.011)\log\sigma - (0.067\pm0.026) ~, \\
\mgtwo & = & (0.199\pm0.016)\log\sigma - (0.168\pm0.038) ~.
\label{eqn:yxfits}
\end{eqnarray}
As expected, the simple regressions yield slopes which are biased low
due to the presence in the data of errors in the abscissa as well as the
ordinate, and also the intrinsic scatter in the relation. Slightly
less-biased results are obtained by least squares regression minimising
the orthogonal residuals (\cf\ J{\o}rgensen \etal\ 1996):
\begin{eqnarray}
\mgbp  & = & (0.109\pm0.012)\log\sigma - (0.078\pm0.027) ~, \\
\mgtwo & = & (0.215\pm0.017)\log\sigma - (0.205\pm0.041) ~.
\label{eqn:orfits}
\end{eqnarray}
These least squares fits and their uncertainties are obtained using the
{\sc slopes} regression program written by E.D.Feigelson and described
in Isobe \etal\ (1990) and Feigelson \& Babu (1992). The uncertainties
are under-estimated because these regressions do not properly account
for the measurement errors or the selection functions. We conclude that
previous determinations of the slope of the \mgsig\ relation are likely
to be biased low whenever the dataset being fitted had significant
errors in the velocity dispersions (as is generally the case). Hereafter
we adopt the ML fits to the \mgsig\ relation.

The distributions of the residuals in \mgbp\ and \mgtwo\ about the ML
fits are shown in the insets to Figures~\ref{fig:mgsig}a
and~\ref{fig:mgsig}b. In order to minimise the effects of outliers, we
robustly estimate the scatter about the \mgsig\ relations as half the
range spanned by the central 68\% of the data points. We find an {\em
observed} scatter of 0.022$\pm$0.002~mag about the \mgbpsig\ relation
and 0.031$\pm$0.003~mag about the \mgtwosig\ relation. Excluding
outliers, the distributions of residuals are very well fitted by
Gaussians parametrised by the median residual and the robustly estimated
scatter. There is no evidence for a tail of negative residuals such as
noted by Burstein \etal\ (1988) and J{\o}rgensen \etal\ (1996). As the
latter authors point out, the presence of such a tail is sensitive to
the adopted slope of the \mgsig\ relation. Some giant ellipticals do,
however, have intrinsically weak Mg linestrengths for their velocity
dispersions (Schweizer \etal\ 1990).

\begin{figure}
\centering
\plotone{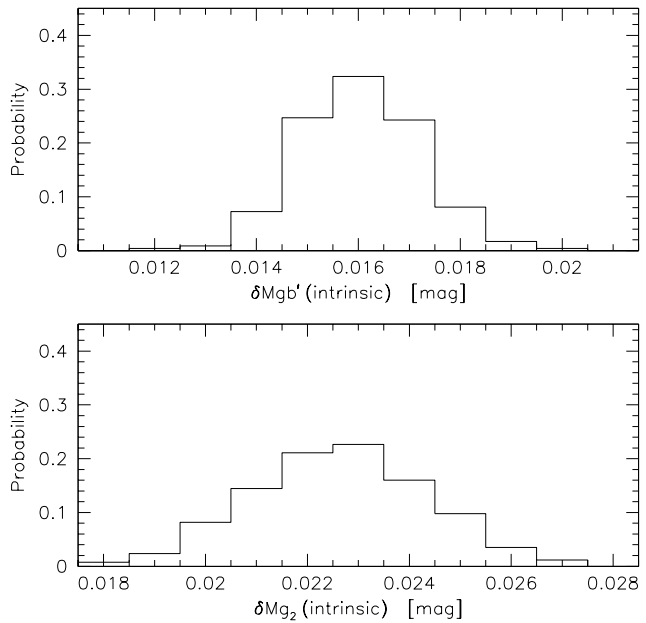}
\caption{The normalised likelihood distributions for the intrinsic
scatter in \mgbp\ and \mgtwo\ at fixed velocity dispersion given the
observed scatter about the \mgbpsig\ and \mgtwosig\ relations. These are
derived from simulations based on the observed distribution of
dispersions and linestrengths (and their errors), assuming only that
there is a global linear \mgsig\ relation about which there is Gaussian
intrinsic scatter.
\label{fig:mgscat}}
\end{figure}

The estimates of the {\em intrinsic} scatter about the relations that
are provided by the ML fit may be exaggerated by outliers or by
deviations of the underlying distribution of galaxies in the \mgsig\
plane from a bivariate Gaussian. We therefore drop the assumption of an
intrinsic bivariate Gaussian distribution in the \mgsig\ plane and use
Monte Carlo simulations based on the observed distribution of
dispersions and linestrengths and their estimated errors (accounting for
both measurement errors and run correction errors). These simulations
assume only that there is a global linear \mgsig\ relation about which
there is Gaussian intrinsic scatter. We vary this intrinsic scatter and
compute the robust estimate of the observed scatter about the fit (the
half-width of the central 68\% of the residuals) for the simulated
distributions. The results of these simulations are presented in
Figure~\ref{fig:mgscat}, which shows the normalised likelihood
distributions for the intrinsic scatter in \mgbp\ and \mgtwo\ given the
observed scatter. We find that to account for the observed scatter in
the relations we require an intrinsic scatter of 0.016$\pm$0.001~mag for
\mgbpsig\ and 0.023$\pm$0.002~mag for \mgtwosig. The ratio of the
intrinsic scatter in \mgtwo\ to the intrinsic scatter in \mgbp\ is
slightly lower than expected from the observed \mgtwo--\mgbp\ relation,
\mgtwo$\approx$1.94\mgbp$-$0.05 (see Paper~2).

\begin{table}
\centering
\caption{Comparison of \mgsig\ relation fits}
\label{tab:mgsig}
\begin{tabular}{lcccc}
 & Slope & Intercept & $\delta$Mg$_{obs}$ & $\delta$Mg$_{int}$ \\
\vspace*{3pt}
{\em \mgbpsig\ relation}\ldots&        &            &            &           \\
Ziegler \& Bender (1997)  & \phpm0.106 &   $-$0.079 &        --- &        ---\\
EFAR (this work)          & \phpm0.131 &   $-$0.131 & \phpm0.022 & \phpm0.016\\
                          & $\pm$0.020 & $\pm$0.048 & $\pm$0.002 & $\pm$0.001\\
\vspace*{3pt}						 
{\em \mgtwosig\ relation}\ldots&       &            &            &           \\
Burstein \etal\ (1988)    & \phpm0.175 &   $-$0.110 & \phpm0.016 & \phpm0.013\\
Guzm\'{a}n \etal\ (1992)  & \phpm0.260 &   $-$0.316 & \phpm0.016 & \phpm0.013\\
                          & $\pm$0.027 & $\pm$0.003 &            &           \\
Bender \etal\ (1993)      & \phpm0.200 &   $-$0.166 & \phpm0.025 & \phpm0.018\\
J{\o}rgensen \etal\ (1996)& \phpm0.196 &   $-$0.155 & \phpm0.025 & \phpm0.020\\
                          & $\pm$0.016 &            &            &           \\
EFAR (this work)          & \phpm0.257 &   $-$0.305 & \phpm0.031 & \phpm0.023\\
                          & $\pm$0.028 & $\pm$0.067 & $\pm$0.003 & $\pm$0.002\\
\end{tabular}
\end{table}

\begin{figure*}
\centering
\plottwo{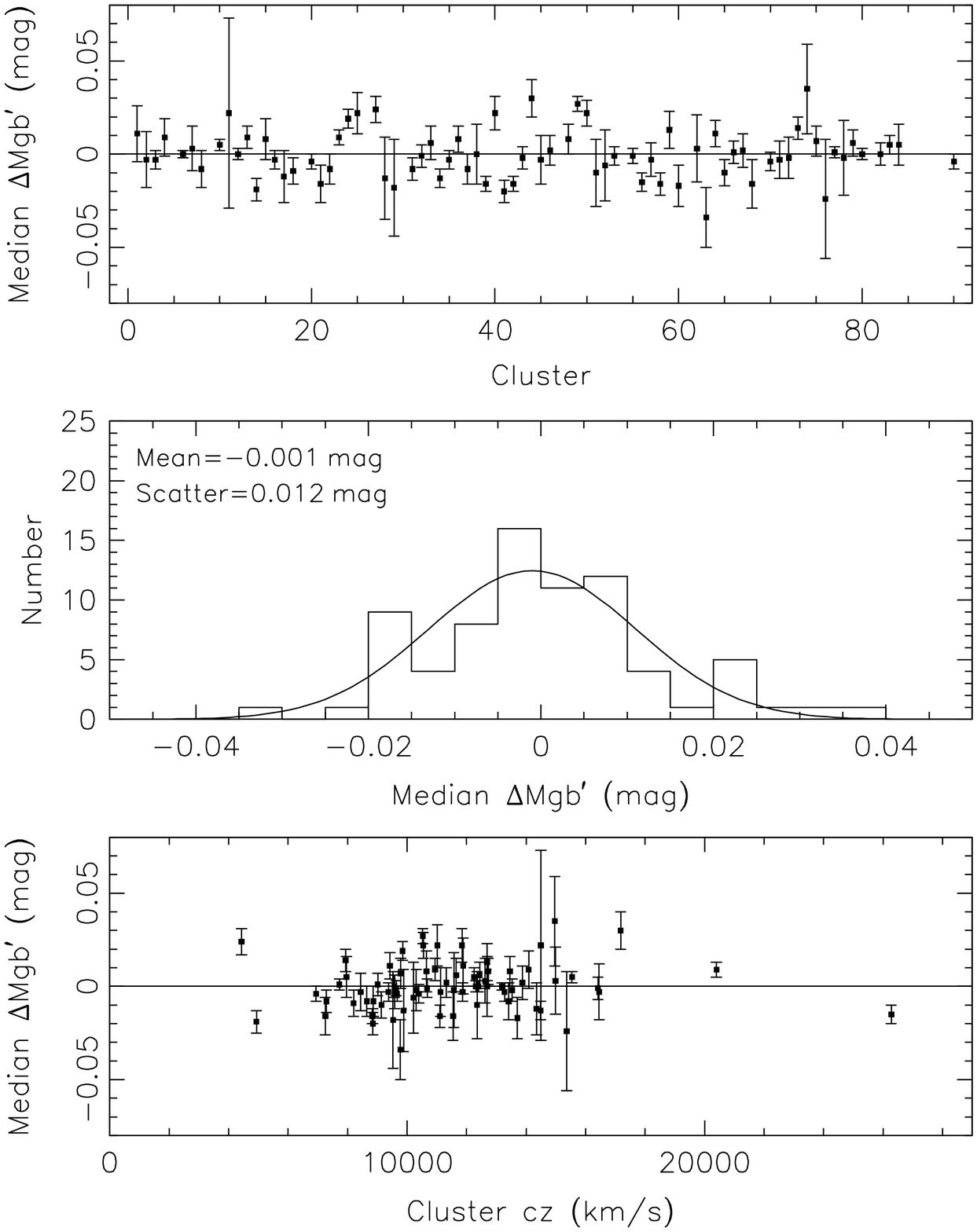}{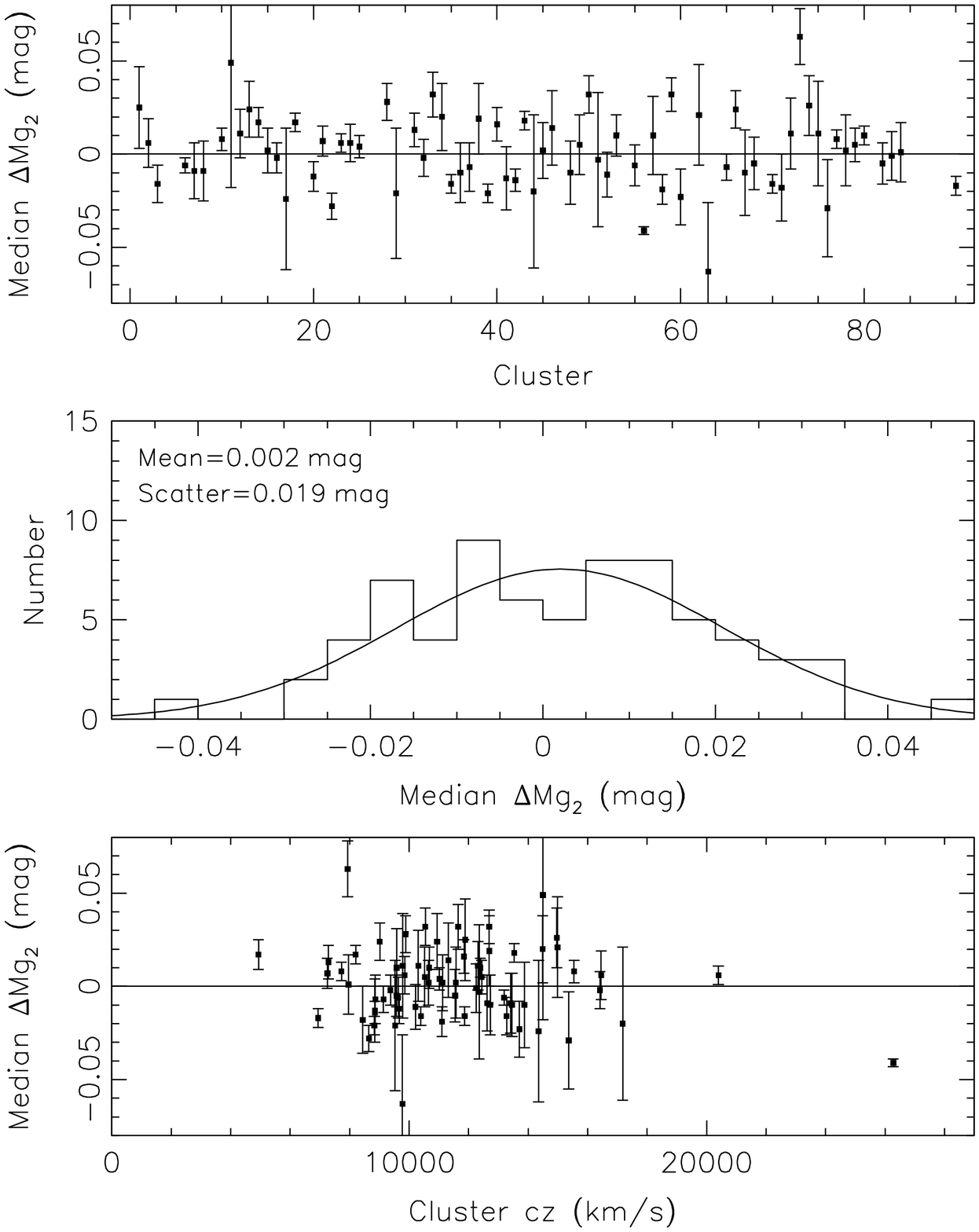}
\caption{Cluster-to-cluster offsets in the \mgsig\ relation. The left
panels are for \mgbpsig\ and the right panels for \mgtwosig. The top
panels show the median offsets from the global \mgsig\ relation for
clusters with three or more linestrength measurements. The middle panels
show the distribution of offsets compared to a Gaussian with the same
mean and dispersion; the scatter is 0.012~mag in \mgbp\ and 0.019~mag in
\mgtwo. The bottom panels show the offsets as a function of redshift.
\label{fig:mgclus}}
\end{figure*}

Table~\ref{tab:mgsig} compares our fits to the \mgsig\ relations
obtained by other authors, and gives the observed scatter
$\delta$Mg$_{obs}$ and the intrinsic scatter $\delta$Mg$_{int}$ in the
relations obtained in each case. For both \mgbpsig\ and \mgtwosig\ the
slopes we obtain are about 25\% steeper than those obtained by most
previous authors. This is not due to a difference in our data, but stems
from our use of the ML method rather than regressions. In this situation
regressions are biased towards flatter slopes than the true relation
because they ignore the intrinsic scatter, the presence of errors in
both variables and the selection function of the dataset. The standard
or orthogonal regression fits to our data, which our simulations show
under-estimate the slope of the relations, give results very similar to
those obtained by other authors.

If we divide the sample by morphological type, we find that the cDs have
a zeropoint which is 0.009~mag higher than that of the other early-type
galaxies in \mgbp, and 0.014~mag higher in \mgtwo. These differences in
the zeropoints are readily apparent in the distributions of residuals
about the global \mgsig\ relations (see the insets to
Figures~\ref{fig:mgsig}a\&b), and are significant at the
3$\sigma$-level. Despite these zeropoint offsets, including or excluding
the cDs changes the scatter about the ML fit by less than its
uncertainty, as they make up only 10\% of the whole sample.

We find no significant differences, however, if we compare the \mgsig\
relations for the two volumes of space, the Hercules-Corona-Borealis and
Perseus-Pisces-Cetus regions, from which our sample is drawn. The two
regions have \mgsig\ relations with slopes and zeropoints which are
consistent both with each other and with the overall \mgsig\ relations,
providing a check that there are no gross systematic environmental
differences between these two regions.

\subsection{Cluster-to-cluster variations}
\label{ssec:clusvars}

We do not have enough galaxies per cluster to fit both the slope and the
zeropoint of the \mgsig\ relations on a cluster-by-cluster basis, even
in our best-sampled clusters. We therefore limit ourselves to
investigating the variation in the \mgsig\ zeropoint. To this end we
measured the median offset in \mgbp\ and \mgtwo\ from the global fits
given above for the clusters with three or more linestrength
measurements (75 clusters for \mgb\ and 72 for \mgtwo). Note that we
only used galaxies that are cluster members based on their redshifts
(see Paper~2). The results are not changed significantly if we use all
clusters, or only clusters with five or more measurements.

\begin{figure*}
\centering
\plottwo{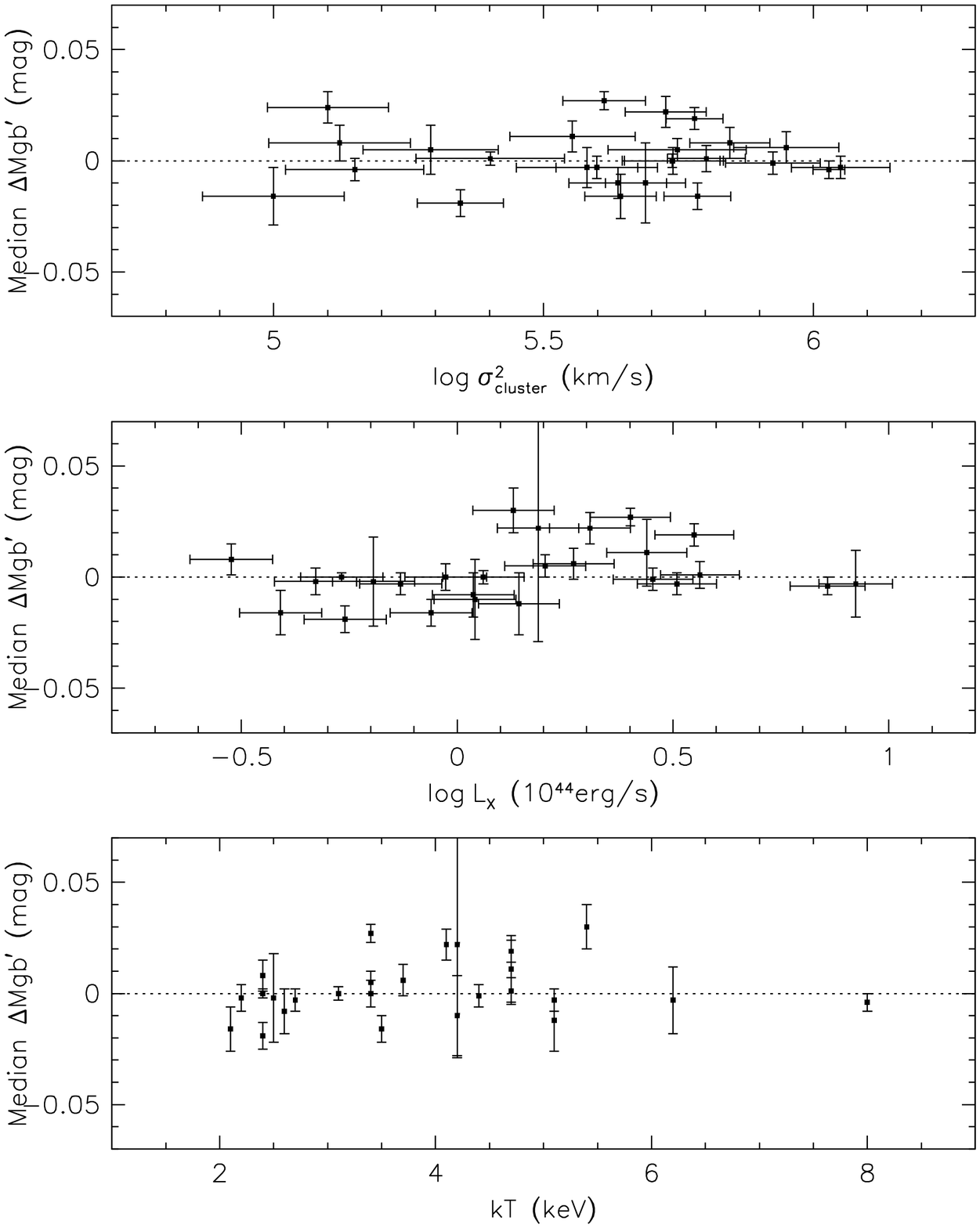}{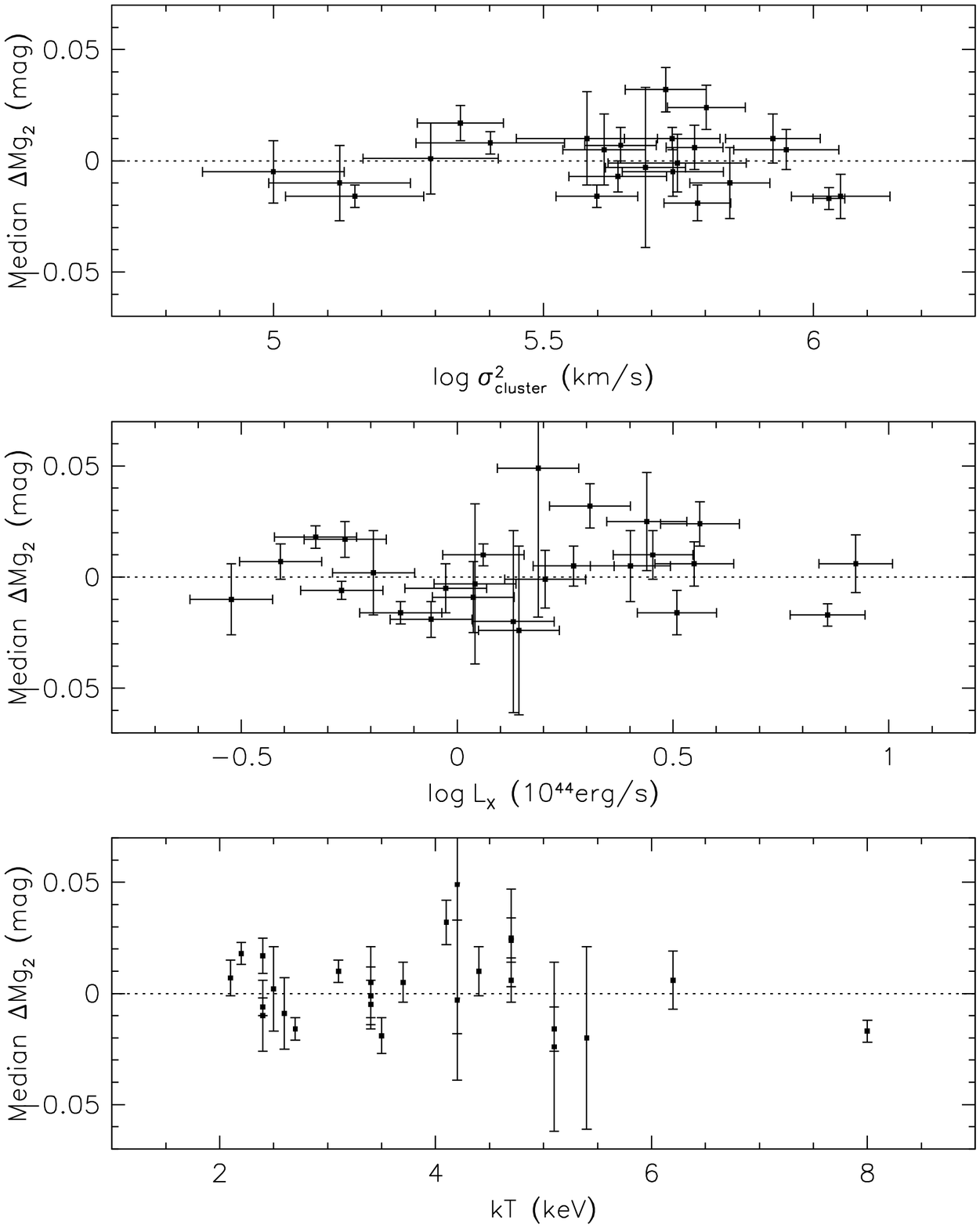}
\caption{Variations in the \mgsig\ relation with various indicators of
cluster mass. The left panels are for \mgbpsig\ and the right panels for
\mgtwosig. The top panels show the median offsets from the global
\mgsig\ relation as a function of $\log\sigma^2_{cluster}$; the middle
panels show the offsets as a function of $\log L_X$; the bottom panels
show the offsets as a function of X-ray temperature, $kT$.
\label{fig:mgrich}}
\end{figure*}

The top panels of Figure~\ref{fig:mgclus} show these zeropoint offsets
as a function of cluster ID number (CID), while the middle panels show
the distributions of the offset values. The robustly-estimated scatter
in the zeropoint offsets is 0.012$\pm$0.002~mag in \mgbpsig\ and
0.019$\pm$0.004~mag in \mgtwosig, showing that the relations are
remarkably uniform among the aggregates of galaxies in the EFAR sample.
The bottom panels in the figure plot the same offsets as a function of
redshift, showing that there is no dependence of the relations on
relative distance within the sample.

This scatter in the zeropoint offsets could purely be a consequence of a
galaxy-to-galaxy scatter in a global \mgsig\ relation, or it could also
require a variation in the zeropoint of the relation from cluster to
cluster. These possibilities were examined by extending the simulations
described in the previous section, adding a further source of scatter to
the \mgsig\ relation in the form of an intrinsic variation between
clusters in the zeropoint of the relation. For simplicity we assume that
this variation also has a Gaussian distribution.

We find that if we make the extreme assumption that there is
cluster-to-cluster scatter but no intrinsic scatter between galaxies
within a cluster, then zeropoint variations between clusters with an rms
of 0.009~mag in \mgbp\ and 0.015~mag in \mgtwo\ are required to recover
the observed cluster-to-cluster scatter. However this model
under-predicts the observed scatter about the global relation, giving
0.017$\pm$0.001~mag for \mgbp\ and 0.025$\pm$0.002~mag for \mgtwo\
compared to the actual values of 0.022$\pm$0.002~mag and
0.031$\pm$0.003~mag. On the other hand, if we assume that there is no
zeropoint variation between clusters, then the intrinsic scatter between
galaxies required to recover the observed scatter in the global relation
(0.016~mag in \mgbp\ and 0.023~mag in \mgtwo; see previous section)
predicts a scatter in the cluster zeropoints of 0.012$\pm$0.001~mag in
\mgbp\ and 0.016$\pm$0.002~mag in \mgtwo, which is consistent with the
observed values of 0.012$\pm$0.002~mag and 0.019$\pm$0.004~mag within
the joint errors.

We conclude that there is no evidence for significant intrinsic
zeropoint variations between clusters, since sampling a galaxy
population drawn from a single global relation with intrinsic scatter
consistent with the observations can account for the zeropoint
differences between our clusters.

\subsection{Variation with cluster properties}
\label{ssec:clusprops}

As there is very little change in the zeropoint of the \mgsig\ relation
from cluster to cluster, it follows that there can be at most only a
weak dependence of the zeropoint on the properties of the clusters. Here
we investigate the effect of cluster properties on the stellar
populations as reflected in the \mgsig\ zeropoints, considering cluster
velocity dispersions, X-ray luminosities and X-ray temperatures (all
indicators of cluster mass). The cluster dispersions come from Table~7
of Paper~2, using redshifts both from EFAR and from the ZCAT catalogue
(Huchra \etal\ 1992; version of 1997 May 29). X-ray luminosities and
temperatures are available for 26 of our 84 clusters in the homogeneous
and flux-limited catalogue of X-ray properties of Abell clusters by
Ebeling \etal\ (1996) based on ROSAT All-Sky Survey data. The X-ray
luminosities are determined to a typical precision of about 20\%. In
order to have comparable precision in the cluster velocity dispersions,
we only use clusters with dispersions computed from at least 20 galaxy
redshifts; this also leaves 26 clusters, 17 of which are in common with
the X-ray subsample.

Figure~\ref{fig:mgrich} shows the offsets in the \mgsig\ relations as
functions of $\log\sigma^2_{cluster}$, $\log L_X$ and $kT$. Applying the
Spearman rank correlation statistic, we find that there is no
significant correlation between the \mgsig\ offsets and any of these
quantities, and thus no evidence for a trend in the zeropoint of the
\mgsig\ relation with cluster mass. Weighted regressions give best-fit
relations and their uncertainties:
\begin{eqnarray}
\Delta\mgbp &=&(-0.002\pm0.009)\log\sigma^2_{cl}+(0.016\pm0.050), 
\label{eqn:clusfitmgb} \\
\Delta\mgtwo&=&(-0.001\pm0.011)\log\sigma^2_{cl}+(0.002\pm0.060).
\label{eqn:clusfitmg2}
\end{eqnarray}
If we take a complementary approach, splitting the clusters into two
subsamples about the median value of $L_X$ and fitting the \mgsig\
relations to the galaxies of the high-$L_X$ and low-$L_X$ clusters
separately, we again find no significant differences in the slopes or
the zeropoints of the fits, which are compatible with the global fits
obtained above.

\section{DISCUSSION}
\label{sec:discussion}

There are at least four main questions which can be addressed using the
above results.

(i) What are the theoretical implications of the lack of correlation
between the mass of a cluster and the zeropoint of the \mgsig\ relation
for cluster galaxies?

(ii) What effect do the stellar population differences implied by the
observed variations in the \mgsig\ relation have on Fundamental Plane
estimates of distances and peculiar velocities?

(iii) What constraint does the intrinsic scatter about the \mgsig\
relation place on the spread in age, metallicity and mass-to-light ratio
amongst early-type galaxies in clusters?

(iv) What further constraints on these quantities result from combining
the scatter about the \mgsig\ relation with the intrinsic scatter about
the Fundamental Plane?

\subsection{\mgsig\ zeropoint and cluster mass}
\label{ssec:mgsigzpt}

The small scatter in the zeropoint of the \mgsig\ relation from cluster
to cluster, and in particular the lack of correlation between the
\mgsig\ zeropoint and the cluster mass, seems to imply that the mass
over-density on Mpc scales in which an early-type galaxy is found has
little connection with its stellar population and star-formation
history.

The variation of the \mgsig\ relation with cluster properties has
previously been studied in a sample of 11 nearby clusters by
J{\o}rgensen \etal\ (1996) and J{\o}rgensen (1997). Following Guzm\'{a}n
\etal\ (1992), these authors look for a trend in \mgtwosig\ offsets with
the `local density' {\em within} clusters. The estimator of local
density used is $\rho_{cluster} = \log\sigma^2_{cluster}-\log R$, where
$R$ is the projected distance of the galaxy from the cluster centre.
Since $R$ is only a lower limit on the galaxy's true distance from the
cluster centre, this is a rather poor estimator of the true local
density. J{\o}rgensen \etal\ find that the residuals in \mgtwo\ show a
weak trend\footnote{The sign of the trend in equation~6 of J{\o}rgensen
(1997) is incorrect; the coefficient should be $+$0.009 (J{\o}rgensen,
priv.comm.)} with local density, $\Delta\mgtwo \propto
0.009\rho_{cluster}$. Since the residuals do {\em not} correlate with
radius within the cluster (see Figure~3 of J{\o}rgensen (1997)), but
{\em do} show a significant correlation with cluster velocity
dispersion, $\Delta\mgtwo \propto 0.02\log\sigma^2_{cluster}$
(least-squares fit to the data in Figure~5 of J{\o}rgensen (1997)), we
would argue that a more straightforward interpretation of their results
is a correlation of \mgtwosig\ zeropoint with total cluster mass rather
than local density.

A correlation of this amplitude is formally consistent at the 2$\sigma$
level with the distribution of \mgtwo\ offsets versus
$\log\sigma^2_{cluster}$ for the EFAR data (see
equation~\ref{eqn:clusfitmg2}); transforming J{\o}rgensen's result via
the \mgtwo--\mgbp\ relation gives a correlation which is consistent at
the 1.4$\sigma$ level with equation~\ref{eqn:clusfitmgb}. We conclude
that any correlation between the \mgsig\ relation zeropoint and the
cluster mass is sufficiently weak (of order $\Delta\mgtwo \propto
0.02\log\sigma^2_{cluster}$ or less) that it is not reliably established
by the existing data, which are consistent with no correlation at all.

Semi-analytic models for the formation of elliptical galaxies, which
previously neglected metallicity effects (see Kauffmann 1996, Baugh
\etal\ 1996), are only now beginning to incorporate chemical enrichment
and successfully reproduce the general form of the observed
colour--magnitude and \mgsig\ relations (Kauffmann \& Charlot 1998). In
consequence, there are as yet no reliable predictions for the variation
of the \mgsig\ relation zeropoint with cluster mass. The limits given
above, together with limits on the difference in \mgsig\ zeropoints for
field and cluster ellipticals (Burstein \etal\ 1990, de Carvalho \&
Djorgovski 1992, J{\o}rgensen 1997), should provide valuable additional
constraints and encourage further development of chemical enrichment
models within a hierarchical framework for galaxy and cluster formation.

\subsection{Systematic effects on FP distances}
\label{ssec:distances}

We now consider the effects on FP distance estimates of systematic
differences in the stellar populations of early-type galaxies from
cluster to cluster. In \S\ref{ssec:clusvars} we found that the observed
cluster-to-cluster variations in the \mgsig\ zeropoint were consistent
with sampling a single global \mgsig\ relation with intrinsic scatter
between galaxies, and did not {\em require} intrinsic scatter between
clusters. Here we turn the question around and ask how much intrinsic
cluster-to-cluster scatter is {\em allowed} by the observations.

From simulations using the model described in \S\ref{ssec:clusvars},
incorporating intrinsic scatter both between galaxies and between
clusters, we find that the maximum cluster-to-cluster scatter allowed
within the 1$\sigma$ uncertainties in the scatter in the global \mgsig\
relation and the cluster zeropoints is approximately 0.005~mag in
\mgbp\ and 0.010~mag in \mgtwo. For our best-fit ML \mgsig\ relations
and a FP given by $R$$\propto$$\sigma^\alpha I^\beta$ with
$\alpha$$\approx$1.27, this level of cluster-to-cluster scatter would
lead to rms errors in FP distance estimates of up to 10\%. These
systematic errors, resulting from differences in the mean stellar
populations between clusters, would apply even to clusters in which the
FP distance errors due to stellar population differences between
galaxies had been made negligible by observing many galaxies in the
cluster.

We emphasise that our results here do not {\em require} any
cluster-to-cluster scatter, but are {\em consistent} with
cluster-to-cluster scatter corresponding to systematic distance errors
between clusters with an rms of up to 10\%. We therefore cannot
determine from the \mgsig\ relation {\em alone} whether systematic
differences in the mean stellar populations between clusters contribute
significantly (or at all) to the errors in FP estimates of distances and
peculiar velocities. A more effective way of testing for such systematic
differences is by directly comparing each cluster's zeropoint offset
from the global \mgsig\ relation to the ratio of its FP and Hubble
distance estimates; this approach will be investigated in a future
paper.

\subsection{Stellar population models}
\label{ssec:models}

\begin{figure}
\centering
\plotone{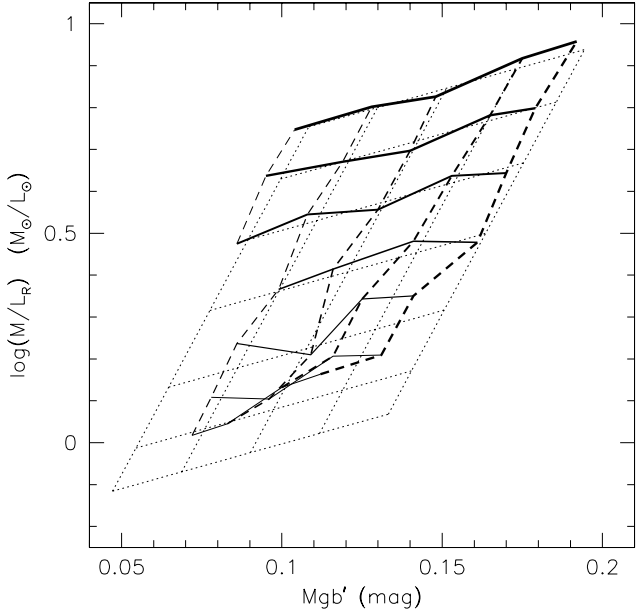}
\caption{The relation between \mgbp\ and $\log M/L_R$ as a function of
age and metallicity in the model of Worthey (1994). The solid lines are
contours of constant age (1.5, 2, 3, 5, 8, 12 and 17~Gyr), with
increasing line thickness indicating increasing age. The dashed lines
are contours of constant metallicity ($-$0.5, $-$0.25, 0, 0.25, 0.5),
with increasing line thickness indicating increasing metallicity. The
dotted grid is the linear fit to the model.
\label{fig:mgbmlrw}}
\end{figure}

To answer the questions concerning the typical age, metallicity and
mass-to-light ratio of early-type galaxies which were raised at the
start of this discussion, we need to employ stellar population models.
We use the predictions from the single stellar population models of
Worthey (1994) and Vazdekis \etal\ (1996), noting the many caveats given
by these authors regarding their models. To simplify our analysis, we
fit \mgbp, \mgtwo\ and $\log M/L_R$ as linear functions of logarithmic
age ($\log t$, with $t$ in Gyr) and metallicity ($\log Z/Z_\odot$), for
galaxies with ages greater than 4~Gyr and metallicities in the range
$-$0.5 to $+$0.5. For the model of Worthey (1994; Salpeter IMF) we
obtain
\begin{eqnarray}
     \mgbp & \approx & 0.058\log t + 0.086\log Z/Z_\odot + 0.077 \\
    \mgtwo & \approx & 0.107\log t + 0.182\log Z/Z_\odot + 0.147 \\
\log M/L_R & \approx & 0.825\log t + 0.184\log Z/Z_\odot - 0.169 ~.
\label{eqn:worthey}
\end{eqnarray}
Figure~\ref{fig:mgbmlrw} compares this fit to Worthey's model in the
case of the predicted dependence of \mgbp\ and $\log M/L_R$ on age and
metallicity. The figure shows that for ages of 5~Gyr or greater the fit
and the model are in satisfactory agreement for all metallicities.

For the model of Vazdekis \etal\ (1996; bimodal IMF, $\mu$=1.35) we have
\begin{eqnarray}
     \mgbp & \approx & 0.051\log t + 0.083\log Z/Z_\odot + 0.084 \\
    \mgtwo & \approx & 0.115\log t + 0.187\log Z/Z_\odot + 0.137 \\
\log M/L_R & \approx & 0.673\log t + 0.251\log Z/Z_\odot - 0.216 ~,
\label{eqn:vazdekis}
\end{eqnarray}
in agreement with the fit obtained by J{\o}rgensen (1997). There is good
agreement between the predictions of the two models for the dependence
of \mgtwo\ and \mgbp\ on age and metallicity, and moderately good
agreement for the dependence of $M/L_R$. 

Note that the same change in the Mg indices is produced by changes in
age, $\Delta\log t$, and metallicity, $\Delta\log Z/Z_\odot$, if
$\Delta\log t/\Delta\log Z/Z_\odot\approx3/2$. This is the `3/2 rule' of
Worthey (1994), which applies to many of the Lick line indices, leaving
them degenerate with respect to variations in age and metallicity.
However age and metallicity produce the same change in $\log M/L_R$ only
if $\Delta\log t/\Delta\log Z/Z_\odot\approx1/3$ or 1/4, so that
measurements of mass-to-light ratios can in principle be combined with
Mg linestrengths to break the age/metallicity degeneracy.

\subsection{Dispersion in age, metallicity and $M/L$}
\label{ssec:scatter}

In the following analysis we infer the dispersion in the ages and
metallicities of early-type galaxies by comparing the scatter in the
\mgsig\ relation with the predictions of the single stellar population
models described in the previous section. This analysis uses the stellar
population models to predict differential changes in the quantities of
interest, and not absolute values. It is also important to remember that
by the dispersion in age or metallicity we mean the dispersion in these
quantities at fixed $\log\sigma$ or, equivalently, the dispersion after
the overall trend with $\log\sigma$ is accounted for. Thus the
dispersion in age or metallicity we infer is the dispersion at fixed
galaxy mass, not the distribution of ages and metallicities as a
function of galaxy mass (which is related to the slope of the \mgsig\
relation and the distribution of galaxies along it).

Single stellar populations models specified by (amongst other
parameters) a unique age and a unique metallicity can only provide an
approximation to real galaxies, whose stellar contents must necessarily
span a range (though perhaps a narrow one) of ages and metallicities.
Since the global Mg indices can be quite sensitive to the detailed
metallicity distribution (Greggio 1997), some of the scatter we observe
may be due to galaxy-to-galaxy differences in the shape of the
metallicity distribution rather than a dispersion in the mean
metallicity or age. 

A further complication is presented by the over-abundance of Mg with
respect to Fe (compared to the solar ratio) in the cores of early-type
galaxies (\eg\ Peletier 1989, Gorgas \etal\ 1990, Worthey \etal\ 1992).
As a comparison of Figures~\ref{fig:mgsig} \&~\ref{fig:mgbmlrw} shows,
the models discussed in the previous section are unable to account for
the highest observed Mg linestrengths. Tantalo \etal\ (1998) have
produced single stellar population models including the effects of
[Mg/Fe] variations and find that
\begin{equation}
\Delta\mgtwo \approx 0.099 \Delta[{\rm Mg}/{\rm Fe}] +
	             0.089 \Delta\log t + 0.166 \Delta\log Z/Z_\odot
\label{eqn:tantalo}
\end{equation}
Comparing this equation with those above, we see that the differential
dependence on age and metallicity is similar to that predicted by
Worthey (1994) and Vazdekis \etal\ (1996). However, any intrinsic
scatter in the [Mg/Fe]--$\sigma$ relation will contribute additionally
to the intrinsic scatter in the \mgtwosig\ relation and reduce the
dispersion in age and metallicity required to account for the
observations.

For these reasons, and also because of other potential sources of
intrinsic scatter such as dark matter, rotation, anisotropy, projection
effects and broken homology, the estimates of the dispersion in age and
metallicity derived here must be considered as upper limits.

With these caveats in mind, we proceed to use the model fits given in
the previous section to infer the dispersion in age or metallicity based
on the observed intrinsic scatter of 0.016~mag in \mgbpsig\ and
0.023~mag in \mgtwosig. For ease of interpretation we quote the
dispersions in age and metallicity as the fractional dispersions $\delta
t/t \equiv \delta\log t/\log e$ and $\delta Z/Z \equiv \delta\log Z/\log
e$. In applying the models in what follows, we adopt the mean of the
coefficients for the two models and give the dispersions in age and
metallicity corresponding to the intrinsic scatter about the \mgbpsig\
relation. Using the intrinsic scatter obtained from the \mgtwosig\
relation would give results that are $\sim$30\% smaller, since the
observed ratio of the intrinsic scatters is
$\delta\mgtwo/\delta\mgbp$$\approx$1.4, rather than about 2 as would be
expected either from the observed \mgtwo--\mgbp\ relation or from the
models. We use the scatter in \mgbp\ rather than \mgtwo\ because our
goal is to establish upper limits on the dispersions in age and
metallicity. The estimated errors in the intrinsic scatter lead to
uncertainties in the dispersions of 5--10\%.

If age variations in single stellar populations are the only source of
scatter then the dispersion in age is $\delta t/t$=67\%, whereas if
metallicity variations are the sole source then the dispersion in
metallicity is $\delta Z/Z$=43\%. Similarly, the observed difference in
the \mgsig\ relation zeropoint for the cD galaxies implies that these
objects are either older or more metal-rich than normal E or E/S0
galaxies. If the zeropoint differences are interpreted as age
differences, cDs are on average 40\% older than typical E or E/S0
galaxies (\ie\ as old as the oldest early-type galaxies); if the
zeropoint differences are interpreted as metallicity differences, cDs
have metallicities on average 25\% higher than typical E or E/S0
galaxies (\ie\ as high as the most metal-rich early-type galaxies).

We can also use the model fits to estimate the approximate change in
$M/L_R$ corresponding to a change in the Mg line indices. If these
changes are caused by age variations alone, then we find that
$\Delta\log M/L_R \approx 7 \Delta\mgtwo$ and $\Delta\log M/L_R \approx
14 \Delta\mgbp$; if, however, they are due only to variations in
metallicity we have $\Delta\log M/L_R \approx 1.2 \Delta\mgtwo$ and
$\Delta\log M/L_R \approx 2.6 \Delta\mgbp$. Thus the change in $\log
M/L_R$ is about 5 times larger if the observed change in the Mg indices
is due to age differences rather than metallicity differences. The
intrinsic scatter in the \mgbpsig\ relation implies a dispersion in
mass-to-light ratio of 50\% if due to age variations, but only 10\% if
due to metallicity variations.

This predicted scatter in $M/L$ is in fact a scatter in luminosity or
surface brightness (since that is all the models deal with). We can
therefore readily establish the effect of this scatter on distances
estimated using the Fundamental Plane (FP) if the scatter in $M/L$ is
uncorrelated with the galaxies' sizes and dispersions, as indeed is the
case for the EFAR sample (at least for galaxies with
$\sigma$$>$100\kms). For a FP given by $R$$\propto$$\sigma^\alpha
I^\beta$, with $R$ the effective radius and $I$ the mean surface
brightness within this radius, if the scatter in $M/L$ is simply a
scatter in $I$ we have $\Delta\log R = \beta \Delta\log M/L$. Most
determinations of the FP, including our own, yield $\beta \approx -0.8$
(\eg\ Dressler \etal\ 1987, J{\o}rgensen \etal\ 1996, Saglia \etal\
1998).

Combining this relation with the dependence of $M/L$ on the Mg line
indices obtained above, we find that the scatter in the \mgsig\ relation
corresponds to an intrinsic scatter in relative distances estimated from
the FP of 40\% if due to age variations, or 8\% if due to metallicity
variations. As the intrinsic scatter in the FP is found to be in the
range 10--20\% (Djorgovski \& Davis 1987, J{\o}rgensen \etal\ 1993,
J{\o}rgensen \etal\ 1996), one cannot explain both the scatter in the
\mgsig\ relation and the scatter in the FP as the result of age
variations alone or metallicity variations alone (unless the single
stellar population models are incorrect or there are significant
galaxy-to-galaxy differences in the metallicity distributions). Suitable
combinations of age variations and metallicity variations {\em can},
however, account for the measured intrinsic scatter in both the \mgsig\
and FP relations.

\subsection{Combined \mgsig\ and FP constraints}
\label{ssec:constraints}

As a simple model, we assume that the scatter in the FP and the \mgsig\
relations (at fixed $\log\sigma$) is entirely due to variations in age
and metallicity (at fixed galaxy mass). These variations are further
assumed to have Gaussian distributions in $\log t$ and $\log Z/Z_\odot$
with dispersions $\delta\log t \equiv \delta t/t\,\log e$ and
$\delta\log Z \equiv \delta Z/Z\,\log e$ and correlation coefficient
$\rho$ ($-$1$\leq$$\rho$$\leq$1). While a Gaussian distribution of
metallicities at fixed galaxy mass is a reasonable initial hypothesis
for describing variations in the chemical enrichment process, the
single-peaked shape of the assumed lognormal distribution for the mean
ages may not realistically represent the star-formation history (even
for galaxies of the same mass). The dispersion in age inferred under
this model should therefore be considered only as a general indication
of the time-span over which early-type galaxies of fixed mass formed the
bulk of their stellar population.

Writing the scatter in Mg linestrengths and FP residuals as $\delta_{\rm
Mg}$ and $\delta_{\rm FP}$ and the dispersion in $\log t$ and $\log
Z/Z_\odot$ as $\sigma_t$ and $\sigma_Z$, this simple model relates the
scatter in the observed quantities to the dispersion in age and
metallicity by:
\begin{eqnarray}
\delta_{\rm Mg}^2 & = & a_t^2 \sigma_t^2 + 2 \rho a_t a_Z \sigma_t \sigma_Z
                      + a_Z^2 \sigma_Z^2 \label{eqn:modelmg} \\
\delta_{\rm FP}^2 & = & b_t^2 \sigma_t^2 + 2 \rho b_t b_Z \sigma_t \sigma_Z
                      + b_Z^2 \sigma_Z^2 \label{eqn:modelfp} ~. 
\end{eqnarray}
Here $a_t$ and $a_Z$ are the coefficients of $\log t$ and $\log
Z/Z_\odot$ for Mg, and $b_t$ and $b_Z$ the coefficients for $\log R =
-0.8\log M/L$, derived from the mean of the linear fits to the two
stellar population models given in \S\ref{ssec:models}.

\begin{figure}
\centering
\plotone{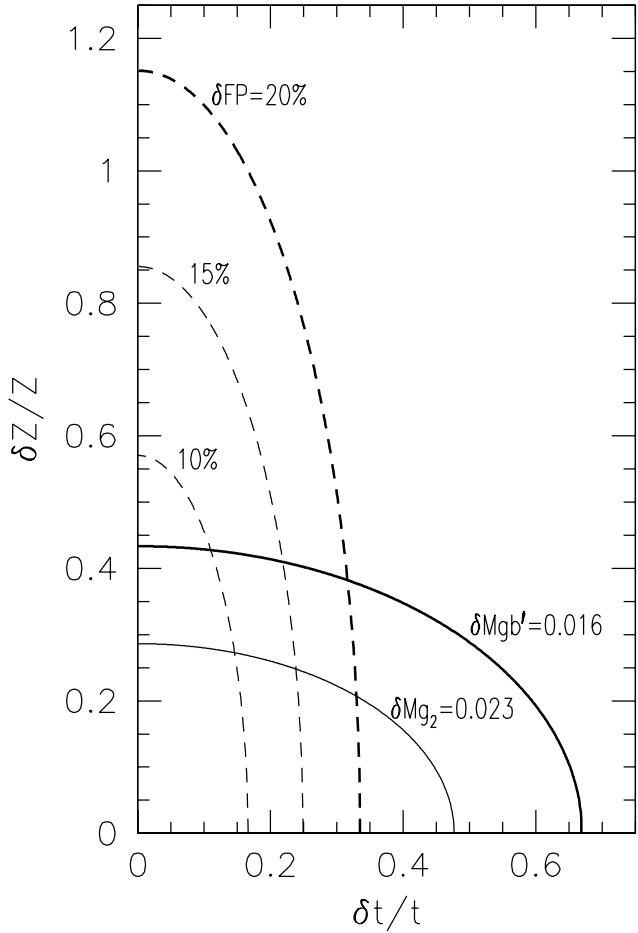}
\caption{Constraints on the dispersion in age, $\delta t/t$, and
metallicity, $\delta Z/Z$, from the intrinsic scatter in the \mgbpsig\
relation ($\delta$\mgbp=0.016~mag; thick solid line) and the \mgtwosig\
relation ($\delta$\mgtwo=0.023~mag; thin solid line), and from an
intrinsic scatter in distance about the FP of 10\%, 15\% and 20\%
(dashed lines).
\label{fig:scatter}}
\end{figure}

\begin{figure}
\centering
\plotone{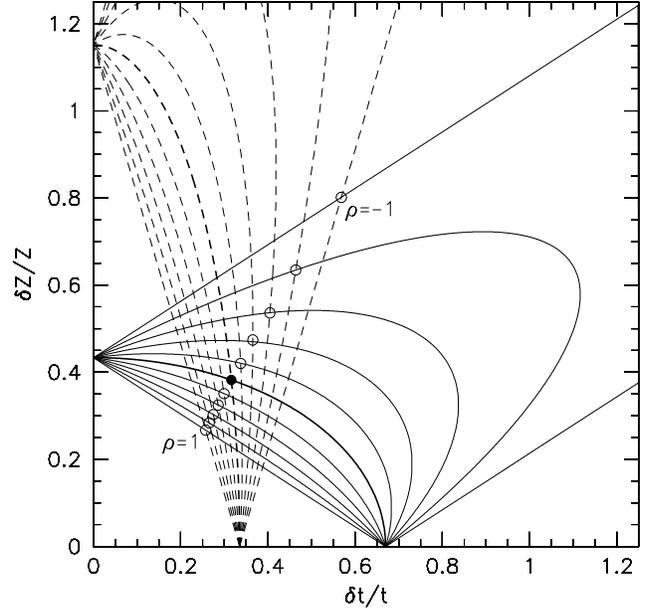}
\caption{Constraints on the dispersion in age, $\delta t/t$, and
metallicity, $\delta Z/Z$, as a function of the correlation coefficient
$\rho$. For values of $\rho$ between $-$1 and 1 (in steps of 0.2) the
constraints imposed by an intrinsic scatter in the \mgbpsig\ relation of
$\delta$\mgbp=0.016~mag are shown by solid lines, while the constraints
from an intrinsic scatter in distance about the FP of 20\% are shown by
dashed lines. The joint constraint solutions for each value of $\rho$
are marked by circles. The uncorrelated case ($\rho$=0) is indicated by
a filled circle and thicker lines.
\label{fig:scatter_rho}}
\end{figure}

Figure~\ref{fig:scatter} shows the constraints on the variations in age
and metallicity (assumed for now to be uncorrelated) which are imposed
by the measured intrinsic scatter in the \mgsig\ relations and the
intrinsic dispersion in $\log M/L_R$ inferred from the intrinsic scatter
in the FP. The intrinsic scatter we find about the \mgbpsig\ and
\mgtwosig\ relations is then consistent with dispersions in age and
metallicity on an elliptical locus defined by equation~\ref{eqn:modelmg}
(with $\rho$=0) in the $\delta t/t$--$\delta Z/Z$ plane. The different
loci for \mgbp\ and \mgtwo\ (the solid lines in
Figure~\ref{fig:scatter}) result from the difference between the
observed ratio of the scatter in \mgtwo\ to that in \mgbp\ and the
predicted ratio from the model, and give some indication of
uncertainties both in the intrinsic scatter about the \mgsig\ relations
and in the model predictions. A second constraint is similarly obtained
from the intrinsic scatter in distance (\ie\ in $\log R$) about the FP
using equation~\ref{eqn:modelfp} (again with $\rho$=0). The dashed lines
in Figure~\ref{fig:scatter} correspond to intrinsic scatter about the FP
of 10\%, 15\% and 20\%.

The important point to note about the figure is that, as mentioned in
\S\ref{ssec:models}, the dependences of the Mg linestrengths and
mass-to-light ratio on age and metallicity are quite different, so that
(if variations in age and metallicity are uncorrelated) the two sets of
constraints are nearly orthogonal. Thus the region of the $\delta
t/t$--$\delta Z/Z$ plane that is consistent with the scatter in both the
\mgsig\ relation and the FP is quite limited. If we use the intrinsic
scatter in the \mgbpsig\ relation and assume a 20\% intrinsic scatter in
$\log R$ about the FP (at the upper end of the quoted range---see, \eg,
Djorgovski \& Davis (1987) or J{\o}rgensen \etal\ (1996)), we obtain
approximate upper limits on the dispersions in age and metallicity of
$\delta t/t$=32\% and $\delta Z/Z$=38\%. If, however, we use the
intrinsic scatter in the \mgtwosig\ relation and adopt an intrinsic FP
scatter of 10\% (as obtained for Coma by J{\o}rgensen \etal\ 1993), then
we obtain approximate lower limits of $\delta t/t$=15\% and $\delta
Z/Z$=27\%.

Similar arguments allow us to evaluate the relative contributions of the
dispersions in age and metallicity to the errors in distance estimates
derived from the FP. For the fiducial case ($\delta$FP=20\%,
$\delta$\mgbp=0.016~mag and $\rho$=0), where $\delta t/t$=32\% and
$\delta Z/Z$=38\%, the mean stellar population model implies that the
dispersion in age gives an intrinsic FP scatter of 19\% while the
dispersion in metallicity gives 7\%. In fact for most of the plausible
range of dispersions in age and metallicity shown in
Figure~\ref{fig:scatter}, it is the dispersion in age which dominates
the intrinsic scatter about the FP. Only for the lowest plausible age
dispersion and the highest plausible metallicity dispersion ($\delta
t/t$=11\% and $\delta Z/Z$=43\%, corresponding to $\delta$FP=10\% and
$\delta$\mgbp=0.016~mag) does the contribution to the FP scatter from
the dispersion in metallicity achieve equality with the contribution
from the dispersion in age.

\begin{figure*}
\centering
\plotfull{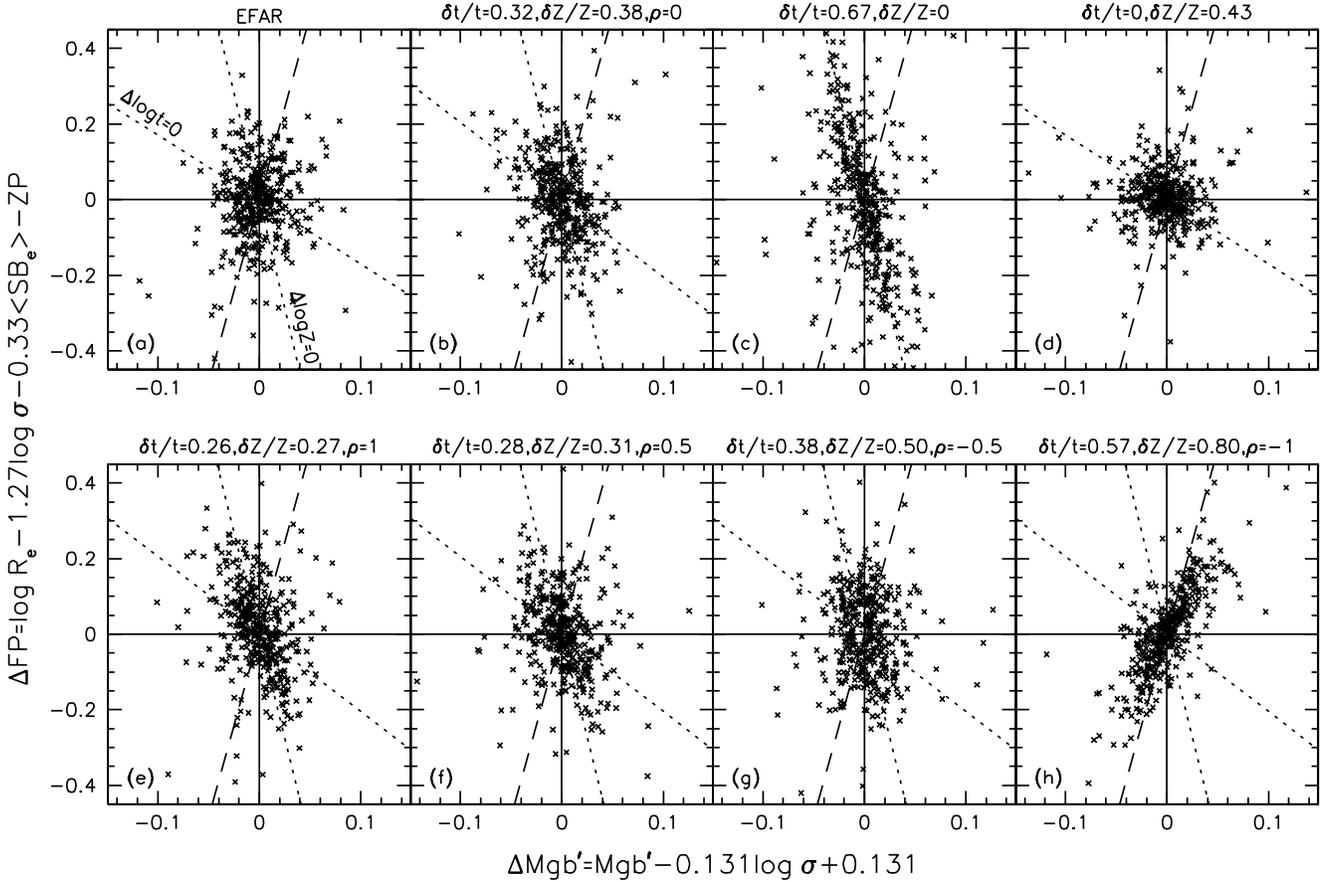}
\caption{The joint distribution of residuals from the \mgbpsig\ and FP
relations: (a)~the residuals for the EFAR data set; (b)~a simulation
with $\delta t/t$=32\% and $\delta Z/Z$=38\% (the values derived from
the intrinsic scatter in the \mgbpsig\ relation and a FP scatter of 20\%
assuming the variations in age and metallicity are uncorrelated); (c)~a
simulation with $\delta t/t$=67\% and $\delta Z/Z$=0 (from the \mgbpsig\
intrinsic scatter assuming a dispersion in age only); (d)~a simulation
with $\delta t/t$=0 and $\delta Z/Z$=43\% (from the \mgbpsig\ intrinsic
scatter assuming a dispersion in metallicity only). The bottom panel
shows simulations consistent with the intrinsic scatter in the \mgbpsig\
relation and a FP scatter of 20\%, and with various assumed correlations
between age and metallicity: (e)~a simulation with $\delta t/t$=26\%,
$\delta Z/Z$=27\% and $\rho$=1; (f)~~a simulation with $\delta
t/t$=28\%, $\delta Z/Z$=31\% and $\rho$=0.5; (g)~a simulation with
$\delta t/t$=38\%, $\delta Z/Z$=50\% and $\rho$=$-$0.5; (h)~a simulation
with $\delta t/t$=57\%, $\delta Z/Z$=80\% and $\rho$=$-$1. The dotted
lines are the expected correlations for a dispersion in age alone
($\Delta\log\,Z$=0) or metallicity alone ($\Delta\log\,t$=0). The dashed
line is the correlation expected if the distribution is dominated by the
errors in $\log\,\sigma$.
\label{fig:dmgdfp}}
\end{figure*}

The constraints on the dispersions change if there is a significant
correlation (or anti-correlation) between the variations in age and
metallicity. Figure~\ref{fig:scatter_rho} shows how the constraints
corresponding to the upper limits $\delta$FP=20\% and
$\delta$\mgbp=0.016~mag (corresponding to the thick lines in
Figure~\ref{fig:scatter}) are modified as the correlation coefficient
$\rho$ varies over its full range from $-$1 to 1. Note that for
$\rho=\pm1$ we have $\delta t/t \propto \mp3/2\,\delta Z/Z$. The main
point to extract from this figure is that if the variations in age and
metallicity have a correlation coefficient in the range
$-$0.5$<$$\rho$$<$1, then the dispersions in age and metallicity vary by
only $\pm$6\% and $\pm$12\% respectively about the values inferred in
the uncorrelated case. Only if the age and metallicity variations are
strongly anti-correlated ($\rho$$\approx$$-$1; \ie\ younger galaxies are
more metal-rich) do we obtain significantly different solutions, with a
broader allowed range in both age and metallicity ($\delta t/t$ as large
as 57\% and $\delta Z/Z$ as large as 80\%). This conclusion is
complementary to that reached by Ferreras \etal\ (1998), who find that
the apparently passive evolution of the colour--magnitude relation
observed in high-redshift clusters does not necessarily imply a common
epoch of major star-formation if younger galaxies are on average more
metal-rich.

We can test the degree of correlation between the variations in age and
metallicity by examining the joint distribution of residuals about the
\mgbpsig\ and FP relations. This distribution is shown for the EFAR data
set in Figure~\ref{fig:dmgdfp}a. There is no evidence for a correlation
between the residuals in this figure; the Spearman rank correlation
coefficient between the residuals is 0.084, and is not significant at
the 2$\sigma$ level. 

In order to investigate the expected distribution of residuals in the
presence of the estimated measurement errors, we have performed Monte
Carlo simulations of the EFAR data using the models for the dispersion
in age and metallicity discussed above. Figure~\ref{fig:dmgdfp}b shows a
simulation with $\delta t/t$=32\% and $\delta Z/Z$=38\%; these are the
values derived from the intrinsic scatter in the \mgbpsig\ relation and
a FP scatter of 20\% when there is no correlation between age and
metallicity. The simulated distribution resembles the observed
distribution, although there is a weak but significant anti-correlation
between the residuals (due to the dominance of the age variations in the
FP residuals) which is not apparent in the EFAR data. Over 100 such
simulations, a two-dimensional K-S test (Press \etal\ 1992) gives a
median probability of 0.3\% that this distribution and the observed
distribution are the same. 

Figures~\ref{fig:dmgdfp}c\&d show simulated distributions for the cases
where the intrinsic scatter in \mgbpsig\ is due to age alone or
metallicity alone. Neither case is consistent with the observed
distribution, supporting the claim that neither age nor metallicity can
be solely responsible for the scatter in both the \mgbpsig\ relation and
the FP. Figures~\ref{fig:dmgdfp}e--h show simulated distributions for
four cases where the variations in age and metallicity are correlated
(with $\rho$=+1, +0.5, $-$0.5 and $-$1 respectively). The perfectly
correlated and perfectly anti-correlated cases are not consistent with
the observed distribution. However Figure~\ref{fig:dmgdfp}g shows that a
distribution with no significant correlation between the \mgbpsig\ and
FP relation residuals is produced when $\rho$=$-$0.5. A two-dimensional
K-S test gives a median probability over 100 such simulations of 1.7\%
that this distribution and the observed distribution are the same. This
relatively low probability may reflect a problem with the model,
although it may simply be due to sampling uncertainty (the probabilities
under this test vary between simulations with an rms of a factor of 6)
or non-Gaussian outliers in the EFAR residuals. The point to be
emphasised is that a model with a moderate degree of anti-correlation
between age and metallicity appears to give significantly better
agreement with the observed distribution than a model in which age and
metallicity are uncorrelated.

\section{CONCLUSIONS}
\label{sec:conclusion}

We have examined the \mgsig\ relation for early-type galaxies in the
EFAR sample. We fit global \mgbpsig\ and \mgtwosig\ relations
(equations~\ref{eqn:mlmgbp} and~\ref{eqn:mlmgtwo}) that have slopes
about 25\% steeper than those obtained by most previous authors. This
difference results not from the data itself but from an improved fitting
procedure: we apply a comprehensive maximum likelihood approach which
correctly accounts for the biases introduced by both the sample
selection function and the significant errors in both Mg and $\sigma$.
The {\em observed} scatter about the \mgsig\ relations is 0.022~mag in
\mgbp\ and 0.031~mag in \mgtwo; the {\em intrinsic} scatter in the
global relations, estimated from Monte Carlo simulations, is 0.016~mag
in \mgbp\ and 0.023~mag in \mgtwo.

With too few galaxies per cluster to reliably determine the full
relation for each cluster separately, we fix the slopes of the relations
at their global values in order to investigate the variation in the
zeropoint from cluster to cluster. We find that the zeropoint has an
observed scatter between clusters of 0.012~mag in \mgbp\ and 0.019~mag
in \mgtwo, and that this observed scatter is consistent with the small
number of galaxies sampled in each cluster being drawn from a single
global relation with intrinsic scatter between galaxies as given
above---\ie\ the observations do not {\em require} any scatter in the
\mgsig\ zeropoint between clusters. The {\em allowed} range for the
intrinsic scatter between clusters corresponds to cluster-to-cluster
systematic errors in Fundamental Plane distances and peculiar velocities
with an rms anywhere in the range 0--10\%. We therefore cannot
determine from the \mgsig\ relation {\em alone} whether systematic
differences in the mean stellar populations between clusters contribute
significantly (or at all) to the errors in distances and peculiar
velocities obtained using the Fundamental Plane.

We have also examined the variation in the \mgsig\ relation with cluster
properties. Our cluster sample ranges from poor clusters to clusters as
rich as Coma, having velocity dispersions from 300\kms\ to 1000\kms\ and
X-ray luminosities spanning 0.3--8$\times$10$^{44}$\,erg\,s$^{-1}$. We
do not detect a significant correlation of \mgsig\ zeropoint with
cluster velocity dispersion, X-ray luminosity or X-ray temperature, nor
is there any significant difference in the \mgsig\ relations obtained by
fitting the galaxies in the high-$L_X$ clusters and low-$L_X$ clusters
separately. The predominant factor in the production of Mg in these
early-type galaxies (and presumably other $\alpha$-elements and perhaps
their metallicity and star-formation history in general) is thus {\em
galaxy} mass and not {\em cluster} mass. These observations place
constraints on semi-analytic models for the formation of elliptical
galaxies, which are now beginning to incorporate chemical enrichment and
should soon be able to make reliable predictions for the variation of
the \mgsig\ relation with cluster mass.

We apply the single stellar population models of Worthey (1994) and
Vazdekis \etal\ (1996) to place upper limits on the global dispersion in
the ages, metallicities and $M/L$ ratios of early-type galaxies of given
mass using the intrinsic scatter in the global \mgsig\ relation. We
infer an upper limit on the dispersion in $M/L_R$ of 50\% if the scatter
in \mgsig\ is due to age differences alone, or 10\% if it is due to
metallicity differences alone. These correspond to upper limits on the
dispersion in relative galaxy distances estimated from the Fundamental
Plane (FP) of 40\% (age alone) or 8\% (metallicity alone). Since the
intrinsic scatter in the FP is found to be 10--20\%, one cannot (within
the context of the single stellar population models) explain both the
scatter in the \mgsig\ relation and the scatter in the FP as the result
of age variations alone or metallicity variations alone.

We therefore determine the joint range of dispersions in age and
metallicity which are consistent with the measured intrinsic scatter in
both the \mgsig\ and FP relations. For a simple model in which the
galaxies have independent Gaussian distributions in $\log t$ and $\log
Z/Z_\odot$, we find upper limits of $\delta t/t$=32\% and $\delta
Z/Z$=38\% at fixed galaxy mass. If the variations in age and metallicity
are not independent, but have correlation coefficient $\rho$, we find
that so long as $\rho$ is in the range $-$0.5 to 1 these limits on the
dispersions in age and metallicity change by only $\pm$6\% and $\pm$12\%
respectively. Only if the age and metallicity variations are strongly
anti-correlated ($\rho$$\approx$$-$1) do we obtain significantly higher
upper limits, with $\delta t/t$ as large as 57\% and $\delta Z/Z$ as
large as 80\%. The distribution of the residuals from the \mgsig\ and FP
relations is only marginally consistent with a model having no
correlation between age and metallicity, and is better-matched by a
model in which age and metallicity variations are moderately
anti-correlated ($\delta t/t$$\approx$40\%, $\delta Z/Z$$\approx$50\%
and $\rho$$\approx$$-$0.5), with younger galaxies being more metal-rich.

Stronger bounds on the dispersion in age and metallicity amongst
early-type galaxies of given mass will require more precise measurements
of the deviations from the \mgsig\ relation and the Fundamental Plane
and also improved models for the dependence of the line indices and
mass-to-light ratio on age and metallicity. Further powerful constraints
can also be obtained by measuring the intrinsic scatter in the \mgsig\
and FP relations at higher redshifts, since the linestrengths and
mass-to-light ratio have different dependences on age.

\section*{Acknowledgements}

MMC acknowledges the support of a DIST Collaborative Research Grant. DB
was partially supported by NSF Grant AST90-16930. RLD thanks the Lorenz
Centre and Prof.\ P.T.~de~Zeeuw. RKM received partial support from NSF
Grant AST90-20864. RPS acknowledges the financial support by the
Deutsche Forschungsgemeinschaft under SFB~375. GW is grateful to the
SERC and Wadham College for a year's stay in Oxford, to the Alexander
von Humboldt-Stiftung for making possible a visit to the
Ruhr-Universit\"{a}t in Bochum and to NSF Grants AST90-17048 and
AST93-47714 for partial support. The entire collaboration benefitted
from NATO Collaborative Research Grant 900159 and from the hospitality
and monetary support of Dartmouth College, Oxford University, the
University of Durham and Arizona State University. Support was also
received from PPARC visitors grants to Oxford and Durham Universities
and PPARC rolling grant `Extragalactic Astronomy and Cosmology in Durham
1994-98'. We thank the referee, Prof.\ Alvio Renzini, for a critique
which resulted in substantial improvements to the paper.

\end{document}